\documentclass[a4paper,fleqn]{cas-sc}

\newcommand{\Ocurrent}{\Omega ( t )}
\newcommand{\Oreference}{\Omega_{0}}
\newcommand{\Ointermediate}{\Omega_{\mathrm{g}}}
\newcommand{\ocurrent}{\omega ( t )}

\newcommand{\bchis}{\boldsymbol{\chi}_{\!\mathrm{s}}}

\newcommand{\bx}{\boldsymbol{x}}
\newcommand{\bX}{\boldsymbol{X}}
\newcommand{\bu}{\boldsymbol{u}}
\newcommand{\bF}{\boldsymbol{F}}
\newcommand{\bFs}{\boldsymbol{F}_{\!\!\mathrm{s}}}
\newcommand{\bFg}{\boldsymbol{F}_{\!\!\mathrm{g}}}
\newcommand{\bFe}{\boldsymbol{F}_{\!\!\mathrm{e}}}
\newcommand{\Js}{J_{\!\mathrm{s}}}
\newcommand{\Jg}{J_{\!\mathrm{g}}}
\newcommand{\Je}{J_{\!\mathrm{e}}}
\newcommand{\bLs}{\boldsymbol{L}_\mathrm{s}}
\newcommand{\bLe}{\boldsymbol{L}_\mathrm{e}}
\newcommand{\bLg}{\boldsymbol{L}_{\mathrm{g}}}

\newcommand{\bv}{\boldsymbol{v}}
\newcommand{\bvs}{\boldsymbol{v}_{\mathrm{s}}}
\newcommand{\bw}{\boldsymbol{w}}
\newcommand{\bd}{\boldsymbol{d}}
\newcommand{\bj}{\boldsymbol{j}}

\newcommand{\bq}{\boldsymbol{q}}

\newcommand{\bsigma}{\boldsymbol{\sigma}}
\newcommand{\bvarsigma}{\boldsymbol{\varsigma}}
\newcommand{\bP}{\boldsymbol{P}}

\newcommand{\bpi}{\boldsymbol{\pi}}
\newcommand{\bC}{\boldsymbol{C}}
\newcommand{\bCe}{\boldsymbol{C}_{\!\mathrm{e}}}

\newcommand{\bM}{\boldsymbol{M}}
\newcommand{\bJ}{\boldsymbol{J}}

\newcommand{\bQ}{\boldsymbol{Q}}
\newcommand{\bI}{\boldsymbol{I}}

\newcommand{\bSigma}{\boldsymbol{\Sigma}}

\newcommand{\bK}{\boldsymbol{K}}

\newcommand{\sumalpha}{\sum_{\alpha = 1}^{N}}
\newcommand{\sumbeta}{\sum_{\beta = 1}^{M}}

\newcommand{\nablax}{\nabla}
\newcommand{\nablaX}{\nabla_{\!0}}

\newcommand{\T}{\mathrm{T}}
\renewcommand{\d}{\mathrm{d}}

\newcommand{\s}{\mathrm{s}}

\newcommand{\exchange}{\mathrm{exch}}
\newcommand{\growth}{\mathrm{growth}}
\newcommand{\internal}{\mathrm{int}}
\newcommand{\external}{\mathrm{ext}}

\newcommand{\bn}{\boldsymbol{n}}
\newcommand{\bN}{\boldsymbol{N}}

\newcommand{\phip}{\phi_{\mathrm{p}}}
\newcommand{\phih}{\phi_{\mathrm{h}}}
\newcommand{\phin}{\phi_{\mathrm{n}}}
\newcommand{\phif}{\phi_{\mathrm{f}}}
\newcommand{\co}{c_{\mathrm{f}}^{\mathrm{o}}}
\newcommand{\cw}{c_{\mathrm{f}}^{\mathrm{w}}}

\newcommand{\tr}{\operatorname{tr}}

\renewcommand{\div}{\operatorname{div}}
\newcommand{\Div}{\operatorname{Div}}

\newcommand{\dt}{\frac{\mathrm{d}}{\mathrm{d} t}}

\usepackage[authoryear]{natbib}

\def\tsc#1{\csdef{#1}{\textsc{\lowercase{#1}}\xspace}}
\tsc{WGM}
\tsc{QE}

\newdefinition{rmk}{Remark}

\begin{document}
\let\WriteBookmarks\relax
\def\floatpagepagefraction{1}
\def\textpagefraction{.001}

\shorttitle{~}    

\shortauthors{J. Stollberg, M.F.P. ten Eikelder and D. Schillinger}  

\title[mode=title]{A thermodynamically consistent framework for finite growth of multi-constituent mixtures with application to tumor growth}

\author[]{Jonathan Stollberg}[orcid=0000-0001-8383-2109]
\cormark[1]
\ead{jonathan.stollberg@tu-darmstadt.de}
\credit{Conceptualization, Methodology, Software, Formal analysis, Investigation, Writing -- Original Draft, Visualization}

\author[]{Marco F.P. {ten Eikelder}}[orcid=0000-0002-1153-146X]
\ead{marco.eikelder@tu-darmstadt.de}
\credit{Methodology, Writing -- Review \& Editing, Supervision}

\author[]{Dominik Schillinger}[orcid=0000-0002-9068-6311]
\ead{dominik.schillinger@tu-darmstadt.de}
\credit{Conceptualization, Writing -- Review \& Editing, Supervision, Project administration, Funding acquisition}

\affiliation[]{organization={Institute for Mechanics, Computational Mechanics Group, Technical University of Darmstadt},
               city={Darmstadt}, 
               postcode={64287}, 
               country={Germany}}

\cortext[1]{Corresponding author}

\begin{abstract}
    Biological tissues grow by continuously producing, transporting, and reorganizing multiple interacting constituents. These processes are intrinsically coupled to finite deformation and residual stress. Existing models typically capture either finite growth kinematics or multi-constituent transport, but rarely both within a thermodynamically consistent setting. In particular, existing approaches do not consistently link the volume created by finite growth to the mass produced for each individual constituent. In this work, we develop a general continuum framework that unifies finite growth kinematics and multiphase mixture theory for fully saturated multi-constituent mixtures containing an arbitrary number of dilute dissolved solutes. Formulated in a solid-skeleton-based description, the framework rests on constituent-wise balance laws and a free-energy dissipation principle, from which thermodynamically admissible constitutive closures are derived for all mass-exchange, transport, reaction, and growth processes. The central novelty of the framework is a coupling between growth-induced volume creation and constituent mass production, expressed through volume accumulation fractions that distribute the newly created volume among the constituents while preserving saturation. We cast the resulting model in a total Lagrangian mixed weak form and specialize the general theory to a four-constituent, two-solute model of avascular tumor growth that couples nutrient transport, waste production, phenotype transitions between proliferative, hypoxic, and necrotic cells, volume growth, elastic deformation, and growth-induced residual stress. The model is implemented within a finite element setting and its capabilities are demonstrated on representative benchmark problems.
\end{abstract}




\maketitle

\section{Introduction}\label{sec:introduction}

Growth is one of the defining features of living matter. Unlike passive engineering materials, biological tissues continuously produce, transport, remove, and reorganize mass. These processes are intrinsically coupled to mechanics: growth changes the local volume and shape of a tissue, mechanical incompatibilities generate elastic deformation and residual stress, and stress can in turn influence transport, proliferation, and remodeling \citep{Taber_1995}. A continuum theory for biological growth must therefore describe not only the evolution of mass, but also the finite deformation and stress generation induced by growth \citep{Goriely_2017}.

A central framework for this purpose is finite growth theory. Following the seminal work of \citet{Rodriguez_1994}, growth is commonly represented by a multiplicative decomposition of the deformation gradient into elastic and growth-related parts, in analogy to multiplicative decompositions in finite inelasticity \citep{Lee_1969, Boyce_1989}. This concept has been developed further in the context of growth-induced residual stresses \citep{Skalak_1996}, balance laws for growing bodies and open-system thermodynamics \citep{Epstein_2000, DiCarlo_2002, Lubarda_2002, Kuhl_2003}, computational growth formulations \citep{Himpel_2005}, stress-modulated growth \citep{Ambrosi_2007}, and growth-induced instabilities \citep{BenAmar_2005, Li_2011}. Finite growth theory has been applied to a wide range of biological systems, including tumor growth \citep{Ambrosi_2002}, cardiac growth \citep{Goektepe_2010, Goektepe_2010b, Rausch_2011}, brain development \citep{Budday_2014, Holland_2015}, skin growth \citep{Buganza_2011}, gastrointestinal organogenesis \citep{Balbi_2015}, and plant morphogenesis involving growing surfaces \citep{Holland_2013}. For broader perspectives on finite tissue growth, we refer to \citep{Taber_1995, Garikipati_2009, Ambrosi_2011, Menzel_2012, Goriely_2017, Ambrosi_2019} and the references therein. These works demonstrate the strength of finite growth kinematics in describing morphogenesis, growth-induced incompatibility, and residual stress. However, many finite growth formulations represent the tissue as an effectively single growing solid and do not explicitly resolve multiple interacting constituents, dissolved species, mass exchange, and chemical reactions.

Many biological tissues are inherently multiphase systems. At the continuum scale, they may be interpreted as mixtures of solid-like cellular and extracellular constituents, interstitial fluid, and dissolved biochemical species \citep{Ambrosi_2011}. Mixture theory and poromechanics provide natural descriptions of such systems. Classical mixture theory describes several constituents occupying the same spatial point and interacting through constituent-wise balance laws and exchange terms \citep{Truesdell_1960, Truesdell_1984}. Poromechanics, originating from Biot's consolidation theory and its finite-deformation extensions \citep{Biot_1941, Biot_1972}, describes deformable fluid-saturated media through the motion of a solid skeleton coupled to pore-fluid transport and pressure-like constraints \citep{Coussy_2003}. A generalization of this concept within mixture theory was introduced as the theory of porous media \citep{Bowen_1980, DeBoer_1990, DeBoer_2000, Ehlers_2001, Ehlers_2002}. Several works have combined growth, transport, and mechanics in mixture-theoretic or poromechanical settings \citep{Humphrey_2002, Garikipati_2004, Ateshian_2007, Ateshian_2010, Ricken_2010, Cyron_2016, Cyron_2017, Humphrey_2021}, with applications to tendon \citep{Garikipati_2004, Narayanan_2009}, regrowth of the resected liver \citep{Ebrahem_2024, Ebrahem_2025}, and tumor growth \citep{Ambrosi_2002a, Byrne_2003, Xue_2016, Mascheroni_2018, Fraldi_2018, Faghihi_2020}. These models provide important ingredients for mechanically coupled tissue growth, but they are typically limited to particular constituent structures, specific transport or reaction mechanisms, or phenomenological growth and exchange laws that are not systematically constrained by thermodynamic principles.

A different route to multiphase biological growth is provided by diffuse-interface models \citep{Cahn_1958, Cahn_1959}, for which general and thermodynamically consistent phase-field formulations for mixtures with an arbitrary number of constituents, so-called $N$-phase mixtures, have recently been developed \citep{Eikelder_2023, Eikelder_2024, Eikelder_2025}. Related approaches have been used to describe nutrient-limited growth, hypoxia, necrosis, evolving interfaces, and interacting species in tumor modeling \citep{Cristini_2003, Wise_2008, Lima_2014, Lorenzo_2016}. However, these diffuse-interface growth models mostly rely on phenomenological closure laws, motivated by biological observations rather than systematically derived or constrained by a common thermodynamic framework. Although their formulations are powerful for multi-constituent transport and phase evolution, they do not naturally incorporate finite deformation effects such as pressure gradients and growth-induced residual stresses, which experiments have shown to play important roles in tumor growth, perfusion, transport, and treatment response \citep{Helmlinger_1997, Roose_2003, Jain_2014, Liu_2020}.

To the best of our knowledge, a general framework that combines finite growth kinematics, multi-constituent mixture theory, and dilute solute transport with thermodynamic admissibility, while consistently coupling growth-induced volume creation to constituent mass production, remains unavailable. Therefore, this paper addresses the need for physically consistent and mathematically rigorous multi-constituent growth models suitable for a wide range of soft-tissue applications. The main contributions and novelties are as follows. First, we unify finite growth kinematics and multiphase mixture theory into a thermodynamically consistent modeling framework applicable to multi-constituent mixtures with an arbitrary number of dilute solute species. Second, we establish a consistent coupling between growth-induced volume creation and constituent mass production, formulated through the notion of volume accumulation fractions that generalize classical single-solid growth closures \citep{Himpel_2005, Buganza_2011} to an arbitrary number of constituents. Within that setup, we derive thermodynamic restrictions and admissible constitutive relations for mass exchange, constituent and solute transport, chemical reactions, and growth. Third, we specialize the general theory to a thermodynamically consistent tumor-growth model that captures mass transport, chemical reactions, phenotype transitions, volume growth, elastic deformation, and the resulting residual stress effects. Taken together, these contributions distinguish the present framework from finite growth models that represent the tissue as a single growing solid, from poromechanical and constrained-mixture growth models restricted to specific constituent structures or phenomenological closures, and from diffuse-interface tumor models that reproduce multi-constituent transport but omit finite growth kinematics and the associated residual stresses.

The remainder of this paper is organized as follows. In Section~\ref{sec:continuum_theory}, we introduce the continuum theory of growing mixtures, including the kinematic description, balance laws, and the coupling between growth kinematics and mass production. Next, in Section~\ref{sec:constitutive_modeling}, we derive the thermodynamic modeling restrictions from a free-energy dissipation law and propose admissible constitutive choices. Subsequently, Section~\ref{sec:implementation} presents the total Lagrangian formulation and the weak form of the coupled mixed problem. Section~\ref{sec:tumor_model} specializes the general theory to a four-constituent, two-solute model for avascular tumor growth, and Section~\ref{sec:results} presents corresponding computational results. Finally, Section~\ref{sec:conclusion} summarizes the main findings and discusses future research avenues.

\section{Continuum theory of growing mixtures}\label{sec:continuum_theory}
In this section, we present our multiphase finite growth framework. It is grounded in continuum mixture theory \citep{Truesdell_1960} and, therefore, obeys the three metaphysical principles formulated in \cite{Truesdell_1984}:
\begin{enumerate}[(i)]
    \item \emph{All properties of the mixture must be mathematical consequences of properties of the constituents.}
    \item \emph{So as to describe the motion of a constituent, we may in imagination isolate it from the rest of the mixture, provided we allow properly for the actions of the other constituents upon it.}
    \item \emph{The motion of the mixture is governed by the same equations as is a single body.}
\end{enumerate}
The first principle states that the mixture is entirely characterized by the properties of its constituents, which are sometimes also referred to as phases. The second principle establishes that the constituents are coupled through interaction terms, which account for their mutual influence. Finally, the third principle asserts that, at the macroscopic level, the motion of the mixture is governed by the same balance laws as those of a single continuum body, rendering it indistinguishable from a single-phase medium in this regard. 

\begin{table}[pos=t]
    \centering
    \caption{Nomenclature of the principal quantities used throughout the general theory, given in the current configuration.}
    \renewcommand{\arraystretch}{1.25}
    \begin{tabular}{ll}
        \toprule
        Symbol & Description \\
        \midrule
        $N$, $M$ & number of constituents and of dissolved solutes \\
        $\alpha, \gamma$; $\beta, \delta$ & constituent and solute indices \\
        $\bI$ & second-order identity tensor \\
        $\Ocurrent$, $\Oreference$, $\Ointermediate$ & current, reference, and intermediate (grown) configuration \\
        $\bx$, $\bX$ & spatial and material position \\
        $\bchis$, $\bu$ & solid-skeleton motion and displacement field \\
        $\bFs$, $\bFe$, $\bFg$ & deformation gradient of the solid skeleton and its elastic and growth parts \\
        $\Js$, $\Je$, $\Jg$ & Jacobians $\det ( \bFs )$, $\det ( \bFe )$, $\det ( \bFg )$ \\
        $\bLs$, $\bLe$, $\bLg$ & spatial, elastic, and growth velocity gradients \\
        $\bvs$, $\bv_{\alpha}$, $\bar{\bv}$ & solid-skeleton, constituent, and barycentric velocity \\
        $\bw_{\alpha}$, $\bd_{\alpha}^{\beta}$ & peculiar fluid velocity and solute diffusion velocity \\
        $\rho_{\alpha}$, $\tilde{\rho}_{\alpha}$, $\rho$ & intrinsic and partial density of constituent $\alpha$, and mixture density \\
        $\phi_{\alpha}$ & volume fraction of constituent $\alpha$ \\
        $c_{\alpha}^{\beta}$, $\tilde{c}_{\alpha}^{\beta}$ & molar concentration of solute $\beta$ dissolved in $\alpha$ per host and per mixture volume \\
        $\mathsf{M}^{\beta}$ & molar mass of solute $\beta$ \\
        $\bj_{\!\alpha}$, $\bj$, $\bq_{\alpha}^{\beta}$ & constituent peculiar, total peculiar, and solute molar diffusive flux \\
        $\zeta_{\alpha}^{\exchange}$, $\zeta_{\alpha}^{\growth}$, $r_{\alpha}^{\beta}$ & inter-constituent mass exchange, mass production, and solute reaction source terms \\
        $\varphi_{\alpha}$, $\theta$ & volume accumulation fraction and its sum over all constituents \\
        $\bpi_{\alpha}$, $\boldsymbol{\vartheta}_{\alpha}$ & linear and angular momentum interaction terms \\
        $\Psi$ & Helmholtz free-energy density per unit current volume \\
        $\mu_{\alpha}$, $\mu_{\lambda,\alpha}$ & chemical potential of constituent $\alpha$ and its saturation-augmented form \\
        $\kappa_{\alpha}^{\beta}$, $\tilde{\kappa}_{\alpha}^{\beta}$ & chemical potential of solute $\beta$ and its volume-fraction-weighted form \\
        $\lambda$ & saturation (Lagrange) multiplier \\
        $\bsigma$, $\bvarsigma$ & Cauchy and Korteweg stress \\
        $\mathcal{A}_{\xi}$, $\mathcal{B}_{\alpha}^{\eta}$ & transition and local reaction affinity \\
        $\mathcal{T}$, $\mathcal{Z}_{\alpha}$ & sets of phenotype transition and solute reaction channels \\
        $\xi$, $\eta$ & labels of phenotype transition channels and solute reaction channels \\
        $\boldsymbol{s}_{\xi}$, $\boldsymbol{\nu}_{\alpha}^{\eta}$ & stoichiometric vectors of phenotype transition and solute reaction channels \\
        $\mathbb{G}$, $\bM_{\alpha\gamma}$, $\bK_{\alpha\gamma}^{\beta\delta}$ & growth, constituent-transport, and solute-diffusion mobility tensors \\
        $m_{\xi}$, $\ell_{\!\alpha\gamma}^{\beta}$, $k_{\alpha}^{\eta}$ & transition, host-switching, and reaction mobilities \\
        $\mathcal{W}$, $\mathcal{D}$ & boundary power and bulk dissipation \\
        \bottomrule
    \end{tabular}%
    \label{tab:notation}%
\end{table}

The notation and fundamental definitions employed in this part of the paper closely follow established formulations in classical mixture theory, as presented in \cite{Eikelder_2023, Eikelder_2024, Eikelder_2025}. At the same time, the kinematic description departs from the fully symmetric mixture-theoretic setting and instead adopts the solid-skeleton-based framework commonly used in nonlinear poromechanics and the theory of porous media \citep{Biot_1972, DeBoer_2000, Ehlers_2002, Coussy_2003} as well as constrained mixture theory \citep{Humphrey_2002, Cyron_2016, Humphrey_2021}. We introduce this kinematic description in the following, beginning with Section~\ref{sec:kinematics}. After defining volume fractions and molar concentrations in Section~\ref{sec:volume_fractions}, we proceed with the formulation of balance laws in Section~\ref{sec:balance_laws}. We finalize our theoretical framework by coupling volume growth and mass source terms in Section~\ref{sec:growth_coupling}. For better readability, we provide a nomenclature of the principal quantities used throughout the general theory in Table~\ref{tab:notation}.

\subsection{Kinematics}\label{sec:kinematics}
\begin{figure}[pos=htbp]
    \centering
    \includegraphics{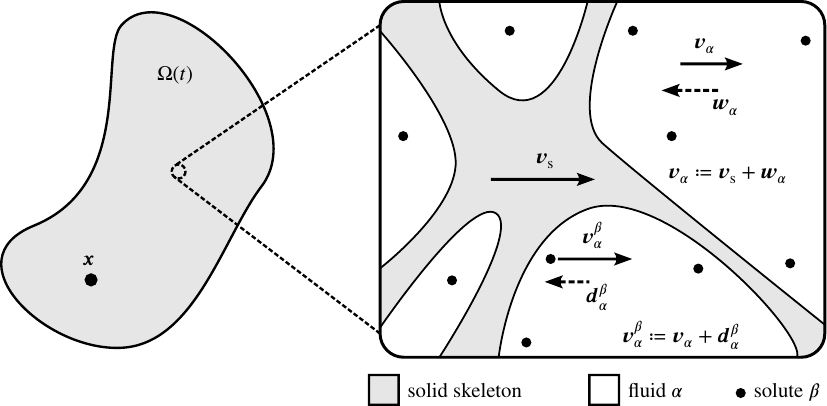}
    \caption{Schematic representation of the transport relative to the solid skeleton. The velocity of pore fluid $\alpha$ is decomposed into the motion of the solid skeleton and its peculiar velocity, whereas the velocity of a solute $\beta$ dissolved in $\alpha$ is decomposed into the motion of the host constituent and the diffusive velocity.}
    \label{fig:solid_skeleton}
\end{figure}

We consider a time-dependent domain $\Ocurrent \subset \mathbb{R}^{d}$ with $d \in \{ 2, 3 \}$ being the spatial dimension. Each spatial point $\bx \in \Ocurrent$ may be simultaneously occupied by an arbitrary number of solid and fluid constituents, indicated by $\alpha = 1,\dots,N$, with dilute dissolved solutes $\beta = 1,\dots,M$. We admit the same maximal number $M$ of solute species in every constituent, so that $M$ does not depend on $\alpha$. A solute that is absent from a given host is represented by a vanishing concentration. The solid constituents of the mixture form a solid skeleton whose pores are filled with the fluid components, see Figure~\ref{fig:solid_skeleton}. 

\begin{figure}[pos=htbp]
    \centering
    \includegraphics{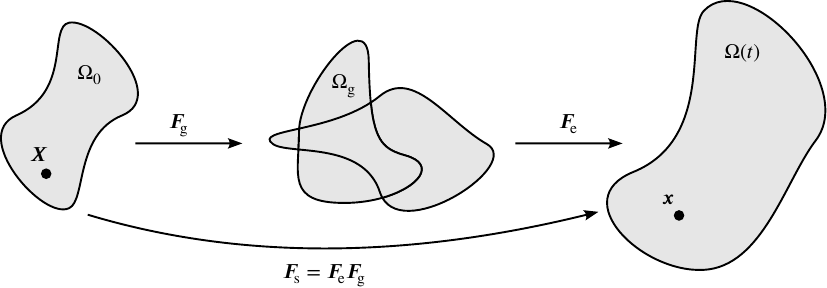}
    \caption{Finite growth kinematics on the mixture level as introduced in \cite{Rodriguez_1994}. Incompatibilities in the displacement field after the growth step are resolved through subsequent elastic deformation.}
    \label{fig:finite_growth}%
\end{figure}

Furthermore, the deformation of the mixture is governed by that of the solid skeleton alone, and we assume the existence of a smooth, invertible map $\bchis : \Oreference \times \mathbb{R}^{+} \to \Ocurrent$ whereby $\Oreference \subset \mathbb{R}^{d}$ denotes the (Lagrangian) reference domain. The spatial position of any material point $\bX \in \Oreference$ at time $t$ is then obtained from
\begin{align}
    \bx \coloneq \bchis ( \bX, t ) \,. 
\end{align}
Accordingly, we define the deformation gradient of the solid skeleton
\begin{align}
    \bFs \coloneq \frac{\partial \bchis}{\partial \bX} = \nablaX \bu + \bI \,,
\end{align}
with the displacement field $\bu \coloneq \bchis - \bX$ and the second-order identity tensor $\bI$. The gradient operator with respect to material coordinates is denoted $\nablaX$. Each spatial point $\bx$ is surrounded by a volume $V \subset \Ocurrent$ with measure $\lvert V \rvert$. Moreover, we introduce the partial volume $V_{\!\alpha} \subset V$ with measure $\lvert V_{\!\alpha} \rvert$ that is occupied by species $\alpha$. Analogously, we can assign each material point $\bX$ a volume $V_{0} \subset \Oreference$ with measure $\lvert V_{0} \rvert$. Note that we assume the solutes to be sufficiently dilute that the volume they occupy is negligible. Within this setup we define the Jacobian
\begin{align}
    \Js \coloneq \lim_{\lvert V_{0} \rvert \to 0} \frac{\lvert V \rvert}{\lvert V_{0} \rvert} = \det ( \bFs ) \,. 
\end{align}
The deformation gradient may be multiplicatively decomposed into an elastic part $\bFe$ and a growth-related part $\bFg$, following the classical finite growth formulation of \cite{Rodriguez_1994}:
\begin{subequations}
    \begin{align}
        \bFs &= \bFe \bFg \,, \label{eq:multiplicative_split} \\
        \Je &\coloneq \det ( \bFe ) \,,\\
        \Jg &\coloneq \det ( \bFg ) \,. 
    \end{align}
\end{subequations}
This decomposition introduces a stress-free intermediate configuration $\Ointermediate$, as illustrated in Figure~\ref{fig:finite_growth}. Growth occurring independently at each material point generally leads to incompatibilities in the displacement field. These incompatibilities are subsequently resolved by an additional elastic deformation, which restores global compatibility and gives rise to residual stresses.

Next, we introduce the material time derivative of an arbitrary differentiable field $\psi$ in the frame shared by the solid mixture constituents
\begin{align}
    \dot{\psi} &\coloneq \frac{\d}{\d t} \psi ( \bX, t ) = \frac{\partial \psi ( \bX, t )}{\partial t} \Big\vert_{\bX} = \frac{\partial \psi ( \bx, t )}{\partial t} \Big\vert_{\bx} + \nablax \psi ( \bx, t ) \cdot \bvs ( \bx, t ) \,,
\end{align}
whereby the velocity $\bvs$ is derived from the deformation map of the solid skeleton through
\begin{align}
    \bvs ( \bx, t ) \coloneq \frac{\d}{\d t} \bchis ( \bX, t ) \,,
\end{align}
and the Nabla operator $\nablax$ indicates the spatial gradient. We postulate that each fluid constituent may move relative to the solid skeleton with its own peculiar velocity $\bw_{\alpha}$ such that
\begin{align}
    \bv_{\alpha} ( \bx, t ) &\coloneq \bvs ( \bx, t ) + \bw_{\alpha} ( \bx, t )\label{eq:fluid_velocity}
\end{align}
is the spatial flow velocity of the respective fluid, see Figure~\ref{fig:solid_skeleton}. Note that $\bw_{\alpha} = \boldsymbol{0}$ and, therefore, $\bv_{\alpha} = \bvs$ for solid constituents. Similarly, we postulate the existence of an independent diffusion velocity $\bd_{\alpha}^{\beta}$ for each solute $\beta$ that describes the motion of the solute relative to its host constituent $\alpha$. The effective transport velocity reads
\begin{align}
    \bv_{\alpha}^{\beta} ( \bx, t ) &\coloneq \bv_{\alpha} ( \bx, t ) + \bd_{\alpha}^{\beta} ( \bx, t ) \,. 
\end{align}

\subsection{Volume fractions and molar concentrations}\label{sec:volume_fractions}
We introduce the intrinsic mass density (per constituent volume) and the partial mass density (per mixture volume) of constituent $\alpha$
\begin{subequations}
    \begin{align}
        \rho_{\alpha} ( \bx, t ) &\coloneq \lim_{\lvert V_{\!\alpha} \rvert \to 0} \frac{m_{\alpha} ( V )}{\lvert V_{\!\alpha} \rvert} \,,\\
        \tilde{\rho}_{\alpha} ( \bx, t ) &\coloneq \lim_{\lvert V \rvert \to 0} \frac{m_{\alpha} ( V )}{\lvert V \rvert} \,,
    \end{align}
    \label{eq:mass_densities}%
\end{subequations}
respectively, where $m_{\alpha} ( V )$ is the mass of the constituent in volume $V$. The mass density of the whole mixture is then obtained from summation over the partial densities
\begin{align}
    \rho ( \bx, t ) \coloneq \sumalpha \tilde{\rho}_{\alpha} ( \bx, t ) \,. 
\end{align}
Furthermore, we define the volume fractions $\phi_{\alpha} \in [ 0, 1 ]$ as
\begin{align}
    \phi_{\alpha} ( \bx, t ) \coloneq \lim_{\lvert V \rvert \to 0} \frac{\lvert V_{\!\alpha} \rvert}{\lvert V \rvert}\label{eq:volume_fractions} \,. 
\end{align}
These are constrained to sum to one, i.e., the mixture is fully saturated and void space is excluded:
\begin{align}
    \sumalpha \phi_{\alpha} - 1 = 0\label{eq:saturation_constraint} \,.
\end{align}
Combining the definitions \eqref{eq:mass_densities} and \eqref{eq:volume_fractions} reveals that $\tilde{\rho}_{\alpha} = \phi_{\alpha} \rho_{\alpha}$. In this work, we restrict our considerations to incompressible constituents, i.e., $\rho_{\alpha} = \mathrm{const.} > 0$.

Analogously, let $n_{\alpha}^{\beta} ( V ) = n^{\beta} ( V_{\!\alpha} )$ be the number of moles of species $\beta$ dissolved in a host constituent $\alpha$. We can then define the molar concentration of $\beta$ per host volume and per mixture volume, respectively, as
\begin{subequations}
    \begin{align}
        c_{\alpha}^{\beta} ( \bx, t ) &\coloneq \lim_{\lvert V_{\!\alpha} \rvert \to 0} \frac{n_{\alpha}^{\beta} ( V )}{\lvert V_{\!\alpha} \rvert} \,,\\
        \tilde{c}_{\alpha}^{\beta} ( \bx, t ) &\coloneq \lim_{\lvert V \rvert \to 0} \frac{n_{\alpha}^{\beta} ( V )}{\lvert V \rvert} \,. 
    \end{align}
\end{subequations}
Together with \eqref{eq:volume_fractions} we obtain the identity $\tilde{c}_{\alpha}^{\beta} = \phi_{\alpha} c_{\alpha}^{\beta}$.

Lastly, we define the peculiar mass and molar diffusive fluxes
\begin{subequations}
    \begin{align}
        \bj_{\!\alpha} &\coloneq \tilde{\rho}_{\alpha} \bw_{\alpha} \,,\\
        \bq_{\alpha}^{\beta} &\coloneq \tilde{c}_{\alpha}^{\beta} \bd_{\alpha}^{\beta} \,,
    \end{align}
    \label{eq:relative_fluxes}%
\end{subequations}
for each mixture constituent and dissolved solute, respectively.

\begin{rmk}[Barycentric velocity]
    In classical mixture theory, peculiar mass fluxes are often defined with respect to the barycentric velocity
    \begin{align}
        \bar{\bv} \coloneq \frac{1}{\rho} \sumalpha \tilde{\rho}_{\alpha} \bv_{\alpha} \,,
    \end{align}
    in which case their sum vanishes identically. With the present choice of the solid skeleton velocity $\bvs$ as reference, however, the total mass flux becomes
    \begin{align}
        \sumalpha \bj_{\!\alpha} = \sumalpha \tilde{\rho}_{\alpha} ( \bv_{\alpha} - \bvs ) = \rho ( \bar{\bv} - \bvs ) \,,
    \end{align}
    and therefore does not vanish in general unless $\bvs = \bar{\bv}$.
\end{rmk}

\subsection{Balance laws}\label{sec:balance_laws}
In compliance with the second metaphysical principle of mixture theory, we describe the motion of each constituent and each dissolved solute through distinct sets of balance laws with appropriate interaction terms. Throughout, we restrict ourselves to isothermal conditions and therefore neglect thermal effects, so that no energy balance is required. Furthermore, we assume that growth, mass exchange, and transport evolve slowly compared with mechanical equilibration. Each constituent is therefore taken to remain in mechanical quasi-equilibrium during the evolution. Consequently, in the absence of body forces, the local balance laws that must hold for all $\bx \in \Ocurrent$ and $t \geq 0$ read
\begin{subequations}
    \begin{align}
        \frac{\partial \tilde{\rho}_{\alpha}}{\partial t} + \div ( \tilde{\rho}_{\alpha} \bv_{\alpha} ) &= \zeta_{\alpha}\label{eq:local_mass_balance_constituent} \,,\\
        \frac{\partial \tilde{c}_{\alpha}^{\beta}}{\partial t} + \div \bigl( \tilde{c}_{\alpha}^{\beta} \bv_{\alpha}^{\beta} \bigr) &= r_{\alpha}^{\beta}\label{eq:local_mass_balance_solute} \,, \\
        \div ( \bsigma_{\alpha} ) &= \bpi_{\alpha}\label{eq:local_linear_momentum_balance} \,,\\
        \bsigma_{\alpha} - \bsigma_{\alpha}^{\T} &= \boldsymbol{\vartheta}_{\alpha}\label{eq:local_angular_momentum_balance} \,,
    \end{align}
\end{subequations}
per $\alpha = 1,\dots,N$ and $\beta = 1,\dots,M$, with $\div ( \bullet )$ denoting the spatial divergence operator. The first equation \eqref{eq:local_mass_balance_constituent} is the constituent mass balance. The interaction term $\zeta_{\alpha}$ accounts for (conserved) mass exchange between the constituents, e.g., due to biological conversion processes, and (non-conserved) mass production due to growth or decline, such as cell mitosis or apoptosis. We therefore split this term
\begin{align}
    \zeta_{\alpha} = \zeta_{\alpha}^{\exchange} + \zeta_{\alpha}^{\growth} \,,\qquad \sumalpha \zeta_{\alpha}^{\exchange} = 0 \,,\qquad \sumalpha \zeta_{\alpha}^{\growth} \in \mathbb{R}\label{eq:conditions_on_mass_source} \,.
\end{align}
The assumption of an open system is motivated by the observation that, at the macroscopic scale under consideration, the detailed biophysical processes governing growth are not explicitly modeled and are instead represented in an effective manner. Equation \eqref{eq:local_mass_balance_solute} denotes the molar concentration balance of solute $\beta$ dissolved in constituent $\alpha$ with the chemical reaction term $r_{\alpha}^{\beta}$. The concentration balance may be transformed into a mass balance law by multiplying it by the molar mass $\mathsf{M}^{\beta} = \mathrm{const.} > 0$. Next, \eqref{eq:local_linear_momentum_balance} represents the balance of linear momentum whereby $\bsigma_{\alpha}$ is the partial Cauchy stress tensor. The interaction term $\bpi_{\alpha}$ accounts for momentum exchange due to drag and friction between different constituents. Lastly, equation \eqref{eq:local_angular_momentum_balance} is the constituent balance of angular momentum with the intrinsic moment density $\boldsymbol{\vartheta}_{\alpha}$. The interaction terms must satisfy the following consistency conditions:
\begin{align}
    \sumalpha \sumbeta \mathsf{M}^{\beta} r_{\alpha}^{\beta} = 0 \,,\qquad \sumalpha \bpi_{\alpha} = \boldsymbol{0} \,,\qquad \sumalpha \boldsymbol{\vartheta}_{\alpha} = \boldsymbol{0}\label{eq:compatibility_interaction_terms} \,.
\end{align}
\begin{rmk}[Vanishing mass exchange in the binary case with non-matching densities]
    Consider a non-growing mixture at rest, such that all velocities $\bv_{\alpha}$ and the mass production terms vanish. The mass balance reduces to
    \begin{subequations}
        \begin{align}
            \frac{\partial \tilde{\rho}_{\alpha}}{\partial t} &= \zeta_{\alpha}^{\exchange} \,,\\
            \frac{\partial \phi_{\alpha}}{\partial t} &= \rho_{\alpha}^{-1} \zeta_{\alpha}^{\exchange} \,,
        \end{align}
    \end{subequations}
    where the second equation follows by dividing the mass balance by the intrinsic density $\rho_{\alpha}$. Summing over all constituents $\alpha$ and using the time derivative of the saturation constraint \eqref{eq:saturation_constraint} gives
    \begin{subequations}
        \begin{align}
            \sumalpha \frac{\partial \tilde{\rho}_{\alpha}}{\partial t} &= \sumalpha \zeta_{\alpha}^{\exchange} = 0 \,,\\
            \sumalpha \frac{\partial \phi_{\alpha}}{\partial t} &= \sumalpha \rho_{\alpha}^{-1} \zeta_{\alpha}^{\exchange} = 0 \,.
        \end{align}
        \label{eq:two_phase_special_case}%
    \end{subequations}
    However, for a two-constituent mixture with distinct intrinsic densities $\rho_{1} \neq \rho_{2}$, the two conditions in \eqref{eq:two_phase_special_case} can be satisfied simultaneously only if
    \begin{align}
        \zeta^{\exchange}_{1} = \zeta^{\exchange}_{2} = 0 \,.
    \end{align}
    Thus, inter-constituent mass exchange is precluded in this limiting case.
\end{rmk}

Since the considered domain is assumed to move with the velocity of the solid skeleton, we rewrite the system of balance laws by making use of the fluxes \eqref{eq:relative_fluxes} such that
\begin{subequations}
    \begin{align}
        \frac{\partial \tilde{\rho}_{\alpha}}{\partial t} + \div ( \tilde{\rho}_{\alpha} \bvs + \bj_{\!\alpha} ) &= \zeta_{\alpha}^{\exchange} + \zeta_{\alpha}^{\growth}\label{eq:mass_balance_solid_frame} \,,\\
        \frac{\partial \tilde{c}_{\alpha}^{\beta}}{\partial t} + \div \bigl( \tilde{c}_{\alpha}^{\beta} \bvs + \tilde{\rho}_{\alpha}^{-1} \tilde{c}_{\alpha}^{\beta} \bj_{\!\alpha} + \bq_{\alpha}^{\beta} \bigr) &= r_{\alpha}^{\beta}\label{eq:concentration_balance_solid_frame} \,, \\ 
        \div ( \bsigma_{\alpha} ) &= \bpi_{\alpha} \,,\\
        \bsigma_{\alpha} - \bsigma_{\alpha}^{\T} &= \boldsymbol{\vartheta}_{\alpha} \,.
    \end{align}\label{eq:balance_laws_solid_frame}%
\end{subequations}
By invoking the incompressibility of the constituents and combining \eqref{eq:mass_balance_solid_frame} with \eqref{eq:concentration_balance_solid_frame}, we then derive the convective forms of the balance laws for the constituent mass and solute concentrations
\begin{subequations}
    \begin{align}
        \rho_{\alpha} \dot{\phi}_{\alpha} &= \zeta_{\alpha}^{\exchange} + \zeta_{\alpha}^{\growth} - \div ( \bj_{\!\alpha} ) - \tilde{\rho}_{\alpha} \div ( \bvs )\label{eq:convective_mass_balance} \,,\\
        \tilde{\rho}_{\alpha} \dot{c}_{\alpha}^{\beta} &= \rho_{\alpha} \bigl( r_{\alpha}^{\beta} - \div \bigl( \bq_{\alpha}^{\beta} \bigr) \bigr) - c_{\alpha}^{\beta} \bigl( \zeta_{\alpha}^{\exchange} + \zeta_{\alpha}^{\growth} \bigr) - \nablax c_{\alpha}^{\beta} \cdot \bj_{\!\alpha}\label{eq:convective_concentration_balance} \,. 
    \end{align}
    \label{eq:convective_balance_laws}%
\end{subequations}

Consistent with the principles of mixture theory, the mixture balance equations follow from summing the balance laws \eqref{eq:balance_laws_solid_frame} over all constituents $\alpha = 1,\dots,N$. We emphasize that, owing to the dilute nature of the solutes, we do not introduce a mixture-level balance law for the molar solute concentrations at this point. We thereby arrive at the following set of equations
\begin{subequations}
    \begin{align}
        \frac{\partial \rho}{\partial t} + \div ( \rho \bvs + \bj ) &= \zeta^{\growth}\label{eq:mixture_mass_balance} \,,\\
        \div ( \bsigma ) &= \boldsymbol{0}\label{eq:mixture_linear_momentum_balance} \,,\\
        \bsigma - \bsigma^{\T} &= \boldsymbol{0}\label{eq:mixture_angular_momentum_balance} \,,
    \end{align}
\end{subequations}
whereby \eqref{eq:mixture_linear_momentum_balance} and \eqref{eq:mixture_angular_momentum_balance} represent the standard forms of the balance laws of linear and angular momentum, respectively. The Cauchy stress is given by
\begin{align}
    \bsigma \coloneq \sumalpha \bsigma_{\alpha} \,.
\end{align}
The total mass of the mixture is in general not conserved, see \eqref{eq:mixture_mass_balance}. Changes in mass arise from both non-conserved fluid transport processes and internal production, which are represented by an effective mass flux and a mixture growth term, respectively defined as
\begin{align}
    \bj \coloneq \sumalpha \bj_{\!\alpha} \,,\qquad \zeta^{\growth} \coloneq \sumalpha \zeta_{\alpha}^{\growth} \,.
\end{align}

\subsection{Coupling of growth kinematics and mass production}\label{sec:growth_coupling}

The coupling developed in this section between the volume created by growth and the mass produced ties the growth kinematics of the solid skeleton to the individual constituent mass balances while preserving full saturation of the mixture, and thereby determines how newly created volume is shared among the constituents. We define the spatial velocity gradient as $\bLs \coloneq \nablax \bvs$, which represents the deformation rate tensor of the solid skeleton, since $\bLs = \dot{\bF}_{\!\!\s}\bFs^{-1}$. In analogy, we introduce the elastic and growth rate tensors as
\begin{subequations}
    \begin{align}
        \bLe &\coloneq \dot{\bF}_{\!\!\mathrm{e}} \bFe^{-1}\label{eq:elastic_deformation_rate} \,,\\
        \bLg &\coloneq \dot{\bF}_{\!\!\mathrm{g}} \bFg^{-1}\label{eq:growth_rate} \,.
    \end{align}
    \label{eq:rate_tensors}
\end{subequations}
These quantities are linked through
\begin{align}
    \bLs = \bLe + \bFe \bLg \bFe^{-1} \,.
    \label{eq:link_between_rate_tensors}
\end{align}
This relation follows directly by differentiating the multiplicative decomposition \eqref{eq:multiplicative_split} and using the definitions of the corresponding rate tensors \eqref{eq:rate_tensors}. Taking the trace of \eqref{eq:link_between_rate_tensors} yields the local relative rate of volume change of the current domain $\Ocurrent$ as the additive superposition of elastic and growth contributions
\begin{align}
    \tr ( \bLs ) = \tr ( \bLe ) + \tr ( \bLg )\label{eq:link_trace_rate_tensors} \,,
\end{align}
where $\tr ( \bLs ) = \div ( \bvs )$. Substitution into the convective form of the constituent mass balance \eqref{eq:convective_mass_balance} yields
\begin{align}
    \rho_{\alpha} \dot{\phi}_{\alpha} = \zeta_{\alpha}^{\exchange} + \zeta_{\alpha}^{\growth} - \div ( \bj_{\!\alpha} ) - \tilde{\rho}_{\alpha} ( \tr ( \bLe ) + \tr ( \bLg ) ) \,. 
\end{align}
Division by the intrinsic density $\rho_{\alpha}$, summation over all constituents, and use of both the saturation constraint \eqref{eq:saturation_constraint} and its material time derivative give
\begin{align}
    \sumalpha \rho_{\alpha}^{-1} \bigl( \zeta_{\alpha}^{\exchange} + \zeta_{\alpha}^{\growth} - \div ( \bj_{\!\alpha} ) \bigr) - \tr ( \bLe ) - \tr ( \bLg ) = 0\label{eq:saturation_constraint_rate_form} \,. 
\end{align}
Thus, \eqref{eq:saturation_constraint_rate_form} is the rate compatibility condition associated with the algebraic constraint \eqref{eq:saturation_constraint}. It ensures that admissible processes preserve full saturation of the mixture by requiring the combined contributions of mass exchange among constituents, growth-induced mass production, and fluid transport to balance the total local rate of volume change.

To model the coupling between growth kinematics and mass production, we now focus on the growth contribution in the above relation. We assume that only a fraction $\theta \in [ 0, 1 ]$ of the volume created by growth is filled by newly produced constituent mass, whereas the remaining fraction $1 - \theta$ initially appears as empty pore space. This motivates the requirement
\begin{align}
    \sumalpha \rho_{\alpha}^{-1} \zeta_{\alpha}^{\growth} - \theta \tr ( \bLg ) = 0\label{eq:growth_mass_production_conditions} \,. 
\end{align}
Since the constituents of a mixture may exhibit different growth capacities, we obtain a constituent-wise realization of this condition by introducing non-negative volume accumulation fractions $\varphi_{\alpha}$. These quantify the fraction of the growth-generated volume that is filled by newly produced mass of constituent $\alpha$ in the intermediate configuration, i.e., immediately after growth and prior to any subsequent transport or elastic accommodation. Accordingly, we require
\begin{align}
    \sumalpha \varphi_{\alpha} = \theta \,,\label{eq:accumulation_sum}
\end{align}
and define the mass production terms as
\begin{align}
    \zeta_{\alpha}^{\growth} = \varphi_{\alpha} \rho_{\alpha} \tr ( \bLg )\label{eq:growth_coupling} \,. 
\end{align}
This constitutive assumption is consistent with the saturation-based compatibility condition \eqref{eq:growth_mass_production_conditions}, which is precisely the growth contribution to the rate form \eqref{eq:saturation_constraint_rate_form} of the saturation constraint. The remaining newly created pore space may instantaneously be compensated by constituent transport or reduced by elastic volume changes, so that constraint \eqref{eq:saturation_constraint} remains satisfied. We furthermore emphasize that the constituent-wise split \eqref{eq:growth_coupling} is a modeling choice rather than a thermodynamic necessity: condition \eqref{eq:growth_mass_production_conditions} fixes only the sum \eqref{eq:accumulation_sum} of the accumulation fractions, so that the individual $\varphi_{\alpha}$ are not uniquely determined and must be prescribed as part of the growth model, informed by the physics of the process, to specify which constituents take up the newly created volume.

\begin{figure}[pos=htbp]
    \centering
    \includegraphics{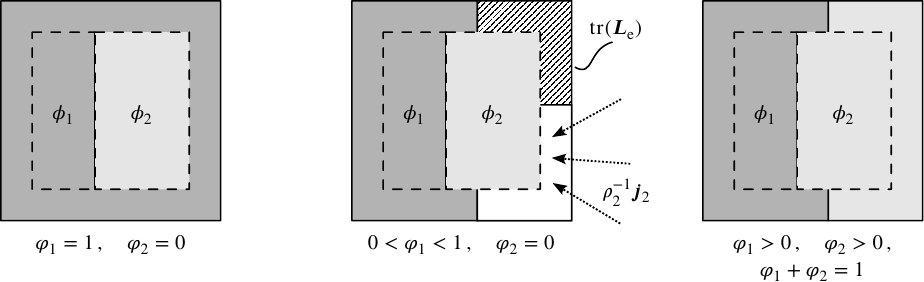}
    \caption{Illustration of the volume-accumulation mechanism underlying the coupling \eqref{eq:growth_coupling}, shown for a two-constituent mixture. Dashed and solid outlines denote a representative volume element before and after the growth increment. Left:~all created volume is accumulated by constituent~1. Center:~constituent~1 accumulates only part of the created volume, so that the remaining pore space is filled by fluid inflow and accommodated by the elastic volume change $\tr ( \bLe )$. Right:~both constituents accumulate mass and jointly fill the created volume.}
    \label{fig:volume_accumulation}%
\end{figure}

This volume-accumulation mechanism is illustrated in Figure~\ref{fig:volume_accumulation} for a two-constituent mixture. In the case of a single-constituent mixture ($N = 1$) where $\varphi_{1} = 1$, this model reduces to the closure proposed by \cite{Himpel_2005} and \cite{Buganza_2011}. For a genuine mixture ($N > 1$), by contrast, the accumulation fractions $\varphi_{\alpha}$ resolve how the created volume is distributed among the individual constituents, an ingredient that, to the best of our knowledge, has not been available in previous finite growth formulations. In this way, the model separates the kinematically induced creation of volume through growth from the constitutive allocation of this volume to constituent mass production.

\section{Constitutive modeling}\label{sec:constitutive_modeling}
The balance laws of Section~\ref{sec:continuum_theory} contain more unknown fields than equations and must therefore be closed by constitutive relations for the fluxes, exchange and reaction rates, growth rate, and stress. We require these relations to be consistent with the second law of thermodynamics, which we enforce through a postulated free-energy dissipation principle. This requirement yields a set of process-wise dissipation inequalities that act as design constraints for every admissible constitutive model. We first state the modeling assumptions in Section~\ref{sec:assumptions}. Based on these assumptions, Section~\ref{sec:restrictions} derives the local dissipation inequality and the resulting thermodynamic restrictions. In Section~\ref{sec:choices}, we then propose admissible constitutive choices that comply with the derived constraints.

\subsection{Modeling assumptions}\label{sec:assumptions}
Since the mixture is described in terms of a single governing deformation field, namely the deformation of the solid skeleton, it is sufficient to consider the reduced set of balance laws
\begin{subequations}
    \begin{align}
        \frac{\partial \tilde{\rho}_{\alpha}}{\partial t}
        + \div ( \tilde{\rho}_{\alpha} \bvs + \bj_{\!\alpha} ) &= \zeta_{\alpha}^{\exchange} + \varphi_{\alpha} \rho_{\alpha} \tr ( \bLg ) \,,\\
        \frac{\partial \tilde{c}_{\alpha}^{\beta}}{\partial t} + \div \bigl( \tilde{c}_{\alpha}^{\beta} \bvs + \tilde{\rho}_{\alpha}^{-1} \tilde{c}_{\alpha}^{\beta} \bj_{\!\alpha} + \bq_{\alpha}^{\beta} \bigr) &= r_{\alpha}^{\beta} \,,\\
        \div ( \bsigma ) &= \boldsymbol{0}
    \end{align}
    \label{eq:eulerian_problem_formulation}%
\end{subequations}
instead of the full set of balance laws \eqref{eq:balance_laws_solid_frame}. The resulting model may be interpreted as a reduced, solid-skeleton-based formulation for a fully saturated constrained mixture, in which constituent transport is described relative to the skeleton, whereas mechanical equilibrium is imposed only at the mixture level \citep{Humphrey_2002}. The unknown fields appearing in this formulation are 
\[
    \phi_{\alpha}\,, \quad
    c_{\alpha}^{\beta}\,, \quad
    \bj_{\!\alpha}\,, \quad
    \bq_{\alpha}^{\beta}\,, \quad
    \zeta_{\alpha}^{\exchange}\,, \quad
    r_{\alpha}^{\beta}\,, \quad
    \bu\,, \quad
    \bLg\,,\quad
    \bsigma
\]
for all $\alpha = 1,\dots,N$ and $\beta = 1,\dots,M$. At this point, we recall that the constituent volume fractions $\phi_{\alpha}$ are subject to the requirement of a fully saturated mixture \eqref{eq:saturation_constraint}. The system is closed by prescribing admissible constitutive relations for the fluxes $\bj_{\!\alpha}$ and $\bq_{\alpha}^{\beta}$, the exchange and reaction rates $\zeta_{\alpha}^{\exchange}$ and $r_{\alpha}^{\beta}$, the growth rate tensor $\bLg$, and the Cauchy stress $\bsigma$. The angular momentum balance does not need to be solved separately. It instead implies the symmetry of the stress tensor and thereby reduces the number of independent components.

To derive restrictions on admissible constitutive choices, we follow the Coleman--Noll procedure \citep{Coleman_1963} in combination with the method of Lagrange multipliers due to \citet{Liu_1972}. To this end, we postulate a mixture-level free-energy dissipation law of the form
\begin{align}
    \frac{\d}{\d t} \int_{\ocurrent} \Psi \,\d v = \mathcal{W} - \mathcal{D} \,,
\end{align}
where $\Psi$ denotes the Helmholtz free-energy density per current mixture volume. The domain $\ocurrent \subset \Ocurrent$ is an arbitrary, time-dependent control volume with volume element $\d v$ that is transported by the velocity of the solid skeleton $\bvs$. We write the boundary of that control volume $\partial \ocurrent$ with surface element $\d a$ and unit outward normal vector $\bn$. The power input through the boundary is collected in $\mathcal{W} = \mathcal{W} ( \partial \ocurrent )$, while $\mathcal{D} = \mathcal{D} ( \ocurrent )$ denotes the total dissipation in the bulk. Thermodynamic admissibility requires
\begin{align}
    \mathcal{D} \geq 0 \,.
\end{align}
Moreover, the free-energy density is assumed to belong to the constitutive class
\begin{align}
    \Psi = \Psi \Bigl( \bFe, \bigl\{ c_{\alpha}^{\beta} \bigr\} , \bigl\{ \phi_{\alpha} \bigr\}, \bigl\{ \nablax \phi_{\alpha} \bigr\} \Bigr)\label{eq:constitutive_energy_class} \,,
\end{align}
where we suppose the elastic deformation gradient $\bFe$ to appear only in the form of dependent objective strain measures. The associated chemical potentials read
\begin{subequations}
    \begin{align}
        \mu_{\alpha} &\coloneq \frac{\partial \Psi}{\partial \phi_{\alpha}} - \div \left( \frac{\partial \Psi}{\partial \nablax \phi_{\alpha}} \right)\label{eq:chemical_potential_phi} \,, \\
        \kappa_{\alpha}^{\beta} &\coloneq \frac{\partial \Psi}{\partial c_{\alpha}^{\beta}}\label{eq:chemical_potential_c} \,.
    \end{align}
    \label{eq:chemical_potential}%
\end{subequations}

In order to expose the thermodynamic structure of the model, we introduce the residuals of the governing equations in local form. From the convective forms of the constituent mass balance and the solute concentration balance \eqref{eq:convective_balance_laws} as well as the balance of linear momentum, we define
\begin{subequations}
    \begin{align}
        \mathcal{M}_{\alpha} &\coloneq \dot{\phi}_{\alpha} - \rho_{\alpha}^{-1} \bigl( \zeta_{\alpha}^{\exchange} - \div ( \bj_{\!\alpha} ) \bigr) + \phi_{\alpha} \tr ( \bLs ) - \varphi_{\alpha} \tr ( \bLg ) \,, \\
        \mathcal{C}_{\alpha}^{\beta} &\coloneq \dot{c}_{\alpha}^{\beta} + \tilde{\rho}_{\alpha}^{-1} \bigl( c_{\alpha}^{\beta} \zeta_{\alpha}^{\exchange} + \nablax c_{\alpha}^{\beta} \cdot \bj_{\!\alpha} \bigr) - \phi_{\alpha}^{-1} \bigl( r_{\alpha}^{\beta} - \div \bigl( \bq_{\alpha}^{\beta} \bigr) \bigr) + \phi_{\alpha}^{-1} c_{\alpha}^{\beta} \varphi_{\alpha} \tr ( \bLg ) \,, \\
        \boldsymbol{\mathcal{P}} &\coloneq \div ( \bsigma ) \,.
    \end{align}
\end{subequations}
In addition, we need to take the residual
\begin{align}
    \mathcal{S} \coloneq - \sumalpha \rho_{\alpha}^{-1} \zeta_{\alpha}^{\exchange} + \sumalpha \rho_{\alpha}^{-1} \div ( \bj_{\!\alpha} ) + \tr ( \bLs ) - \sumalpha \varphi_{\alpha} \tr ( \bLg )\label{eq:saturation_constraint_rate_form_2} \,,
\end{align}
into account, which is equivalent to the rate form \eqref{eq:saturation_constraint_rate_form} of the saturation constraint upon using \eqref{eq:growth_coupling}. We now augment the free-energy dissipation law by these constraints using Lagrange multipliers \citep{Liu_1972}. For the constituent mass balances and solute concentration balances, we choose the respective chemical potentials \eqref{eq:chemical_potential_phi} and \eqref{eq:chemical_potential_c} as multipliers. Moreover, the solid skeleton velocity $\bvs$ is chosen to be the Lagrange multiplier of the balance of linear momentum. This choice is natural since $\bvs = \dot{\bu}$ is the kinematic rate associated with the displacement field and thus the variable entering the mechanical power expenditure of the mixture. Lastly, we introduce an a priori independent scalar multiplier $\lambda$ associated with saturation. This yields the augmented free-energy dissipation law
\begin{align}
    \frac{\d}{\d t} \int_{\ocurrent} \Psi \,\d v + \int_{\ocurrent} \left( \bvs \cdot \boldsymbol{\mathcal{P}} - \sumalpha \mu_{\alpha}\mathcal{M}_{\alpha} - \sumalpha \sumbeta \kappa_{\alpha}^{\beta}\mathcal{C}_{\alpha}^{\beta} - \lambda \mathcal{S} \right) \,\d v = \mathcal{W} - \mathcal{D} \label{eq:augmented_dissipation_inequality_global} \,,
\end{align}
which serves as the thermodynamic setup for the subsequent derivation of local constitutive modeling restrictions.

\subsection{Modeling restrictions}\label{sec:restrictions}
We now derive the local form of the dissipation inequality from the augmented free-energy dissipation law \eqref{eq:augmented_dissipation_inequality_global}. To this end, we first expand the material time derivative of the free-energy functional. By Reynolds' transport theorem, we obtain
\begin{align}
    \frac{\d}{\d t} \int_{\ocurrent} \Psi \,\d v = \int_{\ocurrent} \dot{\Psi} + \Psi \tr ( \bLs ) \,\d v \,. 
\end{align}
Invoking the constitutive dependence \eqref{eq:constitutive_energy_class} together with the definition \eqref{eq:chemical_potential_c}, this expression becomes
\begin{align}
    \frac{\d}{\d t} \int_{\ocurrent} \Psi \,\d v = \int_{\ocurrent} \frac{\partial \Psi}{\partial \bFe} : \dot{\bF}_{\!\!\mathrm{e}} + \sumalpha \sumbeta \kappa_{\alpha}^{\beta} \dot{c}_{\alpha}^{\beta} + \sumalpha \left( \frac{\partial \Psi}{\partial \phi_{\alpha}} \dot{\phi}_{\alpha} + \frac{\partial \Psi}{\partial \nablax \phi_{\alpha}} \cdot \dot{( \nablax \phi_{\alpha} )} \right) + \Psi \tr ( \bLs ) \,\d v \,.
\end{align}
For a scalar field transported with the solid skeleton, the material time derivative of its spatial gradient satisfies
\begin{align}
    \dot{( \nablax \phi_{\alpha} )} = \nablax \dot{\phi}_{\alpha} - \bLs^{\T} \nablax \phi_{\alpha} \,.
\end{align}
Using this identity, integrating the term involving $\nablax \dot{\phi}_{\alpha}$ by parts, and inserting the definition \eqref{eq:chemical_potential_phi}, we obtain
\begin{align}
    \begin{split}
        \frac{\d}{\d t} \int_{\ocurrent} \Psi \,\d v &= \int_{\ocurrent} \frac{\partial \Psi}{\partial \bFe} : \dot{\bF}_{\!\!\mathrm{e}} + \sumalpha \sumbeta \kappa_{\alpha}^{\beta} \dot{c}_{\alpha}^{\beta} + \sumalpha \mu_{\alpha} \dot{\phi}_{\alpha} + \left( \Psi \bI - \sumalpha \nablax \phi_{\alpha} \otimes \frac{\partial \Psi}{\partial \nablax \phi_{\alpha}} \right) : \bLs \,\d v \\
        &\qquad+ \int_{\partial \ocurrent} \sumalpha \dot{\phi}_{\alpha} \frac{\partial \Psi}{\partial \nablax \phi_{\alpha}} \cdot \bn \,\d a \,.
    \end{split}
\end{align}
Together with the kinematic relations \eqref{eq:elastic_deformation_rate} and \eqref{eq:link_between_rate_tensors}, and defining the Korteweg tensor
\begin{align}
    \bvarsigma \coloneq \Psi \bI - \sumalpha \nablax \phi_{\alpha} \otimes \frac{\partial \Psi}{\partial \nablax \phi_{\alpha}} \,,
\end{align}
the mechanical contributions to the free-energy rate may be rewritten as
\begin{align}
        \frac{\partial \Psi}{\partial \bFe} : \dot{\bF}_{\!\!\mathrm{e}} + \bvarsigma : \bLs &= \left( \frac{\partial \Psi}{\partial \bFe} \bFe^{\T} + \bvarsigma \right) : \bLe + \bigl( \bFe^{\T} \bvarsigma \bFe^{-\T} \bigr) : \bLg \,.\label{eq:split_elastic_and_growth_contribution}
\end{align}
Hence, the free-energy rate is written explicitly in terms of the elastic and growth deformation rates and the temporal changes of mixture composition. Next, we insert the residual equations weighted by their respective Lagrange multipliers and perform partial integration of all divergence terms. In this way, boundary contributions are separated from the remaining bulk terms and the individual role of each balance law in the dissipation structure becomes transparent. This gives
\begin{subequations}
    \begin{align}
        \int_{\ocurrent} \bvs \cdot \boldsymbol{\mathcal{P}} \,\d v &= - \int_{\ocurrent} \bsigma : \bLs \,\d v + \int_{\partial \ocurrent} ( \bsigma \bn ) \cdot \bvs \,\d a \,,\\
        \begin{split}
            \int_{\ocurrent} \mu_{\alpha} \mathcal{M}_{\alpha} \,\d v &= \int_{\ocurrent} \mu_{\alpha} \dot{\phi}_{\alpha} - \rho_{\alpha}^{-1} \bigl( \mu_{\alpha} \zeta_{\alpha}^{\exchange} + \nablax \mu_{\alpha} \cdot \bj_{\!\alpha} \bigr) + \phi_{\alpha} \mu_{\alpha} \tr ( \bLs ) - \varphi_{\alpha} \mu_{\alpha} \tr ( \bLg ) \,\d v \\
            &\qquad+ \int_{\partial \ocurrent} \rho_{\alpha}^{-1} \mu_{\alpha} \bj_{\!\alpha} \cdot \bn \,\d a \,,
        \end{split} \\
        \begin{split}
            \int_{\ocurrent} \kappa_{\alpha}^{\beta} \mathcal{C}_{\alpha}^{\beta} \,\d v &= \int_{\ocurrent} \kappa_{\alpha}^{\beta} \dot{c}_{\alpha}^{\beta} + \rho_{\alpha}^{-1} \tilde{\kappa}_{\alpha}^{\beta} \bigl( c_{\alpha}^{\beta} \zeta_{\alpha}^{\exchange} + \nablax c_{\alpha}^{\beta} \cdot \bj_{\!\alpha} \bigr) - \tilde{\kappa}_{\alpha}^{\beta} r_{\alpha}^{\beta} - \nablax \tilde{\kappa}_{\alpha}^{\beta} \cdot \bq_{\alpha}^{\beta} + \varphi_{\alpha} \tilde{\kappa}_{\alpha}^{\beta} c_{\alpha}^{\beta} \tr ( \bLg ) \,\d v \\
            &\qquad+ \int_{\partial \ocurrent} \tilde{\kappa}_{\alpha}^{\beta} \bq_{\alpha}^{\beta} \cdot \bn \,\d a \,,
        \end{split} \\
        \begin{split}
        \int_{\ocurrent} \lambda \mathcal{S} \,\d v &= - \int_{\ocurrent} \sumalpha \rho_{\alpha}^{-1} \lambda \zeta_{\alpha}^{\exchange} + \sumalpha \rho_{\alpha}^{-1} \nablax \lambda \cdot \bj_{\!\alpha} - \lambda \tr ( \bLs ) + \sumalpha \varphi_{\alpha} \lambda \tr ( \bLg ) \,\d v \\
        &\qquad+ \int_{\partial \ocurrent} \sumalpha \rho_{\alpha}^{-1} \lambda \bj_{\!\alpha} \cdot \bn \,\d a \,.
        \end{split}
    \end{align}\label{eq:constraint_terms_energy}%
\end{subequations}
Here, we have introduced the chemical potentials for dissolved solutes weighted by the inverse volume fractions of their respective host constituents,
\begin{align}
    \tilde{\kappa}_{\alpha}^{\beta} \coloneq \phi_{\alpha}^{-1} \kappa_{\alpha}^{\beta} \,,
    \label{eq:weighted_chemical_potential}
\end{align}
which naturally arise from the convective form of the concentration balance. Terms involving the spatial velocity gradient $\bLs$ may be rewritten analogously to \eqref{eq:split_elastic_and_growth_contribution}. Upon collecting the boundary contributions and the remaining bulk terms, the external power and the dissipation are given by
\begin{subequations}
    \begin{align}
        \mathcal{W} &= \int_{\partial \ocurrent} ( \bsigma \bn ) \cdot \bvs + \sumalpha \dot{\phi}_{\alpha} \frac{\partial \Psi}{\partial \nablax \phi_{\alpha}} \cdot \bn - \sumalpha \rho_{\alpha}^{-1} \mu_{\lambda, \alpha} \bj_{\!\alpha} \cdot \bn - \sumalpha \sumbeta \tilde{\kappa}_{\alpha}^{\beta} \bq_{\alpha}^{\beta} \cdot \bn \,\d a \,,\\
        \begin{split}
            \mathcal{D} &= - \int_{\ocurrent} \sumalpha \rho_{\alpha}^{-1} \left( \mu_{\lambda, \alpha} - \sumbeta \tilde{\kappa}_{\alpha}^{\beta} c_{\alpha}^{\beta} \right) \zeta_{\alpha}^{\exchange} + \sumalpha \rho_{\alpha}^{-1} \left( \nablax \mu_{\lambda, \alpha} - \sumbeta \tilde{\kappa}_{\alpha}^{\beta} \nablax c_{\alpha}^{\beta} \right) \cdot \bj_{\!\alpha} \\
            &\qquad  + \sumalpha \sumbeta \tilde{\kappa}_{\alpha}^{\beta} r_{\alpha}^{\beta} + \sumalpha \sumbeta \nablax \tilde{\kappa}_{\alpha}^{\beta} \cdot \bq_{\alpha}^{\beta} + \left( \frac{\partial \Psi}{\partial \bFe} \bFe^{\T} + \bvarsigma - \sumalpha \phi_{\alpha} \mu_{\lambda, \alpha} \bI - \bsigma \right) : \bLe \\
            &\qquad+ \left( \bFe^{\T} ( \bvarsigma - \bsigma ) \bFe^{-\T} - \sumalpha \phi_{\alpha} \mu_{\lambda,\alpha} \bI + \sumalpha \varphi_{\alpha} \left( \mu_{\lambda,\alpha} - \sumbeta \tilde{\kappa}_{\alpha}^{\beta} c_{\alpha}^{\beta} \right) \bI \right) : \bLg
            \,\d v \,,
        \end{split}
    \end{align}
\end{subequations}
where we defined
\begin{align}
    \mu_{\lambda,\alpha} \coloneq \mu_{\alpha} + \lambda \,,
\end{align}
since the chemical potentials of the constituents appear only in combination with the Lagrange multiplier $\lambda$. For a brief discussion of the interconnection between chemical potentials and the Lagrange multiplier enforcing a fully saturated mixture, we refer to \cite{Eikelder_2025}. The boundary power $\mathcal{W}$ then includes traction power and the energetic contributions associated with constituent and solute transport. The corresponding bulk contribution $\mathcal{D}$ contains the products of generalized driving forces and thermodynamic fluxes.

Adopting the Coleman--Noll argument, purely elastic processes are assumed to be non-dissipative. Consequently, the driving force conjugate to $\bLe$ must vanish identically. This yields the constitutive expression for the Cauchy stress:
\begin{align}
    \bsigma = \frac{\partial \Psi}{\partial \bFe} \bFe^{\T} + \bvarsigma - \sumalpha \phi_{\alpha} \mu_{\lambda,\alpha} \bI\label{eq:cauchy_stress} \,.
\end{align}
The stress response is therefore not determined by elastic deformation alone. Rather, the composition of the mixture enters explicitly through the chemical pressure contributions $\phi_{\alpha} \mu_{\alpha}$, which modify the isotropic part of the Cauchy stress together with the pressure-like multiplier $\lambda$. To satisfy the balance of angular momentum \eqref{eq:mixture_angular_momentum_balance}, the Korteweg stress must be symmetric, i.e., $\bvarsigma = \bvarsigma^{\T}$. A sufficient condition is
\begin{align}
    \nablax \phi_{\alpha} \otimes \frac{\partial \Psi}{\partial \nablax \phi_{\alpha}} = \frac{\partial \Psi}{\partial \nablax \phi_{\alpha}} \otimes \nablax \phi_{\alpha}\label{eq:gradient_energy_requirement} \,.
\end{align}
Since the subdomain $\ocurrent$ may be chosen arbitrarily, the dissipation inequality must hold in local form. After inserting the stress relation \eqref{eq:cauchy_stress}, we arrive at
\begin{align}
    \begin{split}
        &- \sumalpha \rho_{\alpha}^{-1} \left( \mu_{\lambda,\alpha} - \sumbeta \tilde{\kappa}_{\alpha}^{\beta} c_{\alpha}^{\beta} \right) \zeta_{\alpha}^{\exchange} - \sumalpha \rho_{\alpha}^{-1} \left( \nablax \mu_{\lambda,\alpha} - \sumbeta \tilde{\kappa}_{\alpha}^{\beta} \nablax c_{\alpha}^{\beta} \right) \cdot \bj_{\!\alpha} \\
        &- \sumalpha \sumbeta \tilde{\kappa}_{\alpha}^{\beta} r_{\alpha}^{\beta} - \sumalpha \sumbeta \nablax \tilde{\kappa}_{\alpha}^{\beta} \cdot \bq_{\alpha}^{\beta} - \left( \sumalpha \varphi_{\alpha} \left( \mu_{\lambda,\alpha} - \sumbeta \tilde{\kappa}_{\alpha}^{\beta} c_{\alpha}^{\beta} \right) \bI  - \bFe^{\T} \frac{\partial \Psi}{\partial \bFe} \right) : \bLg \geq 0 \,.
    \end{split}\label{eq:dissipation_inequality}
\end{align}
\begin{rmk}[Alternative free-energy formulation]
    An equivalent result may also be obtained by formulating the dissipation principle in terms of a Helmholtz free-energy density directly augmented by the saturation constraint \eqref{eq:saturation_constraint}. To this end, we define the augmented free-energy density
    \begin{align}
        \hat{\Psi} \coloneq \Psi \Bigl( \bFe, \bigl\{ c_{\alpha}^{\beta} \bigr\}, \bigl\{ \phi_{\alpha} \bigr\}, \bigl\{ \nablax \phi_{\alpha} \bigr\} \Bigr) + \lambda \left( \sumalpha \phi_{\alpha} - 1 \right) \,,
    \end{align}
    where the additional term does not alter the physical free-energy itself. The chemical potentials include the Lagrange multiplier $\lambda$ a priori since
    \begin{align}
        \frac{\partial \hat{\Psi}}{\partial \phi_{\alpha}} = \frac{\partial \Psi}{\partial \phi_{\alpha}} + \lambda \,.
    \end{align}
    In this alternative formulation, the free-energy dissipation law may be postulated as
    \begin{align}
        \frac{\d}{\d t} \int_{\ocurrent} \hat{\Psi} \,\d v + \int_{\ocurrent} \bvs \cdot \boldsymbol{\mathcal{P}} - \sumalpha \mu_{\alpha} \mathcal{M}_{\alpha} - \sumalpha \sumbeta \kappa_{\alpha}^{\beta} \mathcal{C}_{\alpha}^{\beta} \,\d v = \mathcal{W} - \mathcal{D} \,.
    \end{align}
    Proceeding from this augmented free-energy dissipation law likewise leads to the Cauchy stress expression \eqref{eq:cauchy_stress} and dissipation inequality \eqref{eq:dissipation_inequality}.
\end{rmk}

The dissipation inequality consists of contributions associated with growth, constituent transport, inter-constituent mass exchange, solute transport, and solute reactions. While thermodynamic admissibility requires only the total sum of these terms to be non-negative, our objective is not to construct the most general constitutive theory. Rather, we aim to introduce a coherent set of models that is directly compatible with \eqref{eq:constitutive_constraints}. In this spirit, we choose the constitutive responses such that each individual process contributes non-negatively to the dissipation on its own:
\begin{subequations}
    \begin{align}
        - \sumalpha \rho_{\alpha}^{-1} \left( \mu_{\lambda, \alpha} - \sumbeta \tilde{\kappa}_{\alpha}^{\beta} c_{\alpha}^{\beta} \right) \zeta_{\alpha}^{\exchange} &\geq 0\label{eq:constraint_constituent_exchange} \,,\\
        - \sumalpha \rho_{\alpha}^{-1} \left( \nablax \mu_{\lambda, \alpha} - \sumbeta \tilde{\kappa}_{\alpha}^{\beta} \nablax c_{\alpha}^{\beta} \right) \cdot \bj_{\!\alpha} &\geq 0\label{eq:constraint_constituent_flux} \,,\\
        - \sumalpha \sumbeta \tilde{\kappa}_{\alpha}^{\beta} r_{\alpha}^{\beta} &\geq 0\label{eq:constraint_concentration_exchange} \,,\\
        - \sumalpha \sumbeta \nablax \tilde{\kappa}_{\alpha}^{\beta} \cdot \bq_{\alpha}^{\beta} &\geq 0\label{eq:constraint_concentration_flux} \,,\\
        - \left( \sumalpha \varphi_{\alpha} \left( \mu_{\lambda, \alpha} - \sumbeta \tilde{\kappa}_{\alpha}^{\beta} c_{\alpha}^{\beta} \right) \bI  - \bFe^{\T} \frac{\partial \Psi}{\partial \bFe} \right) : \bLg &\geq 0\label{eq:constraint_growth_rate} \,.
    \end{align}
    \label{eq:constitutive_constraints}%
\end{subequations}

\subsection{Admissible constitutive choices}\label{sec:choices}
Guided by the modeling restrictions derived above, we now specify a set of constitutive laws for the individual dissipative mechanisms. We restrict attention to constitutive laws of the classes
\begin{subequations}
    \begin{align}
        \zeta_{\alpha}^{\exchange} &= \zeta_{\alpha}^{\exchange} \Bigl( \bigl\{ \phi_{\alpha} \bigr\}, \bigl\{ c_{\alpha}^{\beta} \bigr\}, \bigl\{ \mu_{\lambda, \alpha} \bigr\}, \bigl\{ \tilde{\kappa}_{\alpha}^{\beta} \bigr\} \Bigr)\label{eq:class_constituent_exchange} \,,\\
        \bj_{\!\alpha} &= \bj_{\!\alpha} \Bigl( \bigl\{ \phi_{\alpha} \bigr\}, \bigl\{ c_{\alpha}^{\beta} \bigr\}, \bigl\{ \tilde{\kappa}_{\alpha}^{\beta} \bigr\}, \bigl\{ \nablax c_{\alpha}^{\beta} \bigr\}, \bigl\{ \nablax \mu_{\lambda, \alpha} \bigr\} \Bigr)\label{eq:class_constituent_flux} \,,\\
        r_{\alpha}^{\beta} &= r_{\alpha}^{\beta} \Bigl( \bigl\{ \phi_{\alpha} \bigr\}, \bigl\{ c_{\alpha}^{\beta} \bigr\}, \bigl\{ \tilde{\kappa}_{\alpha}^{\beta} \bigr\} \Bigr)\label{eq:class_concentration_exchange} \,,\\
        \bq_{\alpha}^{\beta} &= \bq_{\alpha}^{\beta} \Bigl( \bigl\{ \phi_{\alpha} \bigr\}, \bigl\{ c_{\alpha}^{\beta} \bigr\}, \bigl\{ \nablax \tilde{\kappa}_{\alpha}^{\beta} \bigr\} \Bigr)\label{eq:class_concentration_flux} \,,\\
        \bLg &= \bLg \Bigl( \bFe, \bFg, \bigl\{ \phi_{\alpha} \bigr\}, \bigl\{ c_{\alpha}^{\beta} \bigr\}, \bigl\{ \mu_{\lambda, \alpha} \bigr\}, \bigl\{ \tilde{\kappa}_{\alpha}^{\beta} \bigr\} \Bigr)\label{eq:class_growth_rate} \,.
    \end{align}
\end{subequations}
These constitutive classes directly reflect the structure of the thermodynamic driving forces identified in \eqref{eq:constitutive_constraints} and serve as the point of departure for the specific constitutive choices introduced in the following. In particular, Onsager reciprocal relations play a central role in our modeling framework \citep{Onsager_1931a, Onsager_1931b}. For details on the thermodynamics of chemical reactions and mass transition processes, in particular the role of the affinity, we refer to \cite{Donder_1936}.

\paragraph{Inter-constituent mass exchange.}
Inter-constituent exchange between host constituents is modeled as a chain of phenotype transition processes. To this end, we introduce a set $\mathcal{T}$ of admissible transition channels. Each channel $\xi \in \mathcal{T}$ is characterized by its stoichiometric exchange vector
\begin{align}
    \boldsymbol{s}_{\xi} \coloneq ( s_{1,\xi},\dots,s_{N,\xi} )^{\T} \in \mathbb{R}^{N} \,,\label{eq:stoichiometric_vector}
\end{align}
where negative and positive entries identify the constituents removed and produced by the process, respectively. We then postulate
\begin{align}
    \zeta_{\alpha}^{\exchange} = \sum_{\xi\in\mathcal{T}} s_{\alpha,\xi} \hat{\zeta}_{\xi}^{\exchange}\label{eq:exchange_transition_decomposition}\,,
\end{align}
with stoichiometric coefficients $s_{\alpha,\xi} \in \boldsymbol{s}_{\xi}$ and non-negative phenotype transition rates
\begin{align}
    \hat{\zeta}_{\xi}^{\exchange} = \hat{\zeta}_{\xi}^{\exchange} \Bigl( \bigl\{ \phi_{\alpha} \bigr\}, \bigl\{ c_{\alpha}^{\beta} \bigr\}, \bigl\{ \mu_{\lambda, \alpha} \bigr\}, \bigl\{ \tilde{\kappa}_{\alpha}^{\beta} \bigr\} \Bigr) \,.
\end{align}
Conservation of the total exchanged mass \eqref{eq:conditions_on_mass_source} requires
\begin{align}
    \sumalpha s_{\alpha,\xi} = 0 \qquad \forall\, \xi \in \mathcal{T} \,.
\end{align}
Substituting the channel decomposition \eqref{eq:exchange_transition_decomposition} into the exchange dissipation inequality \eqref{eq:constraint_constituent_exchange} identifies the corresponding transition affinity
\begin{align}
    \mathcal{A}_{\xi} \coloneq - \sumalpha s_{\alpha,\xi} \rho_{\alpha}^{-1}  \left( \mu_{\lambda, \alpha} - \sumbeta \tilde{\kappa}_{\alpha}^{\beta} c_{\alpha}^{\beta} \right) \,,
\end{align}
as the thermodynamic driving force conjugate to $\hat{\zeta}_{\xi}^{\exchange}$, so that a positive affinity marks a thermodynamically favorable transition. The transition rate is then chosen as the unilateral kinetic law
\begin{align}
    \hat{\zeta}_{\xi}^{\exchange} = m_{\xi} \langle \mathcal{A}_{\xi} \rangle_{\!+} \,.\label{eq:transition_rate}
\end{align}
Hence, each transition channel is activated only if its affinity is positive, so that undesired reverse transitions may be excluded. The mobility $m_{\xi} \geq 0$ is assumed to have the same dependencies as the constitutive class \eqref{eq:class_constituent_exchange} and to vanish whenever the volume fraction of any constituent consumed by the transition tends to zero, so that a transition cannot proceed in the absence of its reactants.

\paragraph{Constituent peculiar fluxes.}
For the constituent peculiar mass fluxes, we select Onsager-type relations of the form
\begin{align}
    \bj_{\!\alpha} = - \sum_{\gamma=1}^{N} \bM_{\alpha\gamma} \rho_{\gamma}^{-1} \left( \nablax \mu_{\lambda, \gamma} - \sum_{\beta=1}^{M} \tilde{\kappa}_{\gamma}^{\beta} \nablax c_{\gamma}^{\beta} \right) \,,
\end{align}
where the mobility tensors satisfy the Onsager reciprocal symmetry $\bM_{\alpha\gamma} = \bM_{\gamma\alpha}^{\T}$. In addition, their collection forms a positive semi-definite block operator. More precisely, we require
\begin{align}
    \sumalpha \sum_{\gamma=1}^{N} \boldsymbol{z}_{\alpha}^{\T} \bM_{\alpha\gamma} \boldsymbol{z}_{\gamma} \geq 0 \qquad \forall\, \boldsymbol{z}_{1},\dots,\boldsymbol{z}_{N} \in \mathbb{R}^{d} \,.
\end{align}
With this constitutive law, the constituent peculiar mass fluxes are driven by gradients of the associated thermodynamic potentials together with correction terms arising from dissolved solutes. We assume the mobility tensors to depend on the same state variables as in \eqref{eq:class_constituent_flux} and to be degenerate in the sense that they vanish as any of the involved volume fractions approaches zero or one. Furthermore, whenever $\alpha$ refers to a solid constituent, $\bM_{\alpha \gamma}$ must vanish in order to remain compatible with $\bv_{\alpha} = \bvs$ for the solid skeleton. 

\paragraph{Solute source terms.}
The source terms of the dissolved solutes are decomposed into host-switching and chemical reaction contributions according to
\begin{align}
    r_{\alpha}^{\beta} = \sum_{\gamma=1}^{N}  \hat{r}^{\mathrm{sw},\beta}_{\alpha \gamma} + \sum_{\eta \in \mathcal{Z}_{\alpha}} \nu_{\alpha}^{\beta,\eta} \hat{r}_{\alpha}^{\mathrm{reac}, \eta}\label{eq:split_reaction_source}
\end{align}
with the rates having the same dependencies as \eqref{eq:class_concentration_exchange}, i.e.,
\begin{align}
    \hat{r}^{\mathrm{sw},\beta}_{\alpha \gamma} = \hat{r}^{\mathrm{sw},\beta}_{\alpha \gamma} \Bigl( \bigl\{ \phi_{\alpha} \bigr\}, \bigl\{ c_{\alpha}^{\beta} \bigr\}, \bigl\{ \tilde{\kappa}_{\alpha}^{\beta} \bigr\} \Bigr) \,,\qquad \hat{r}_{\alpha}^{\mathrm{reac},\eta} = \hat{r}_{\alpha}^{\mathrm{reac},\eta} \Bigl( \bigl\{ \phi_{\alpha} \bigr\}, \bigl\{ c_{\alpha}^{\beta} \bigr\}, \bigl\{ \tilde{\kappa}_{\alpha}^{\beta} \bigr\} \Bigr) \,.
\end{align}
The first term describes switching of solute $\beta$ between different host constituents $\alpha$ and $\gamma$. For this purpose, we propose a linear exchange law of the form
\begin{align}
    \hat{r}^{\mathrm{sw},\beta}_{\alpha \gamma} = \ell_{\!\alpha \gamma}^{\beta} \bigl( \tilde{\kappa}_{\gamma}^{\beta} - \tilde{\kappa}_{\alpha}^{\beta} \bigr) \,.\label{eq:solute_host_switching}
\end{align}
To ensure non-negative dissipation and conservation of this transfer process, we require $\ell_{\!\alpha\gamma}^{\beta} = \ell_{\!\gamma\alpha}^{\beta} \geq 0$ (see Appendix~\ref{sec:host_switching_symmetry} for details). The second term represents chemical reactions occurring locally within host constituent $\alpha$, where $\mathcal{Z}_{\alpha}$ denotes the corresponding set of reaction channels. The stoichiometric vector for each host constituent and reaction channel reads
\begin{align}
    \boldsymbol{\nu}_{\alpha}^{\eta} \coloneq \bigl( \nu_{\alpha}^{1,\eta},\dots,\nu_{\alpha}^{M,\eta} \bigr)^{\T} \,,
\end{align}
with its components satisfying
\begin{align}
    \sum_{\beta=1}^{M} \mathsf{M}^{\beta}\nu_{\alpha}^{\beta,\eta} = 0 \qquad \forall\, \alpha \in \{ 1, \dots, N \}\,,\ \forall\, \eta \in \mathcal{Z}_{\alpha} \,,
\end{align}
to enforce mass-conservative reactions, see \eqref{eq:compatibility_interaction_terms}. Here, the superscript $\eta$ labels the reaction channel and $\nu_{\alpha}^{\beta,\eta} \in \boldsymbol{\nu}_{\alpha}^{\eta}$ are the stoichiometric coefficients of the solutes $\beta$ in that reaction, i.e., they are negative for reactants and positive for products. We recall that $\mathsf{M}^{\beta}$ denotes the molar mass of solute $\beta$. Inserting \eqref{eq:split_reaction_source} into the dissipation inequality for solute reactions \eqref{eq:constraint_concentration_exchange} reveals that the associated host-local reaction affinity is given by
\begin{align}
    \mathcal{B}_{\alpha}^{\eta} \coloneq - \sum_{\beta=1}^{M} \nu_{\alpha}^{\beta,\eta} \tilde{\kappa}_{\alpha}^{\beta} \,.
\end{align}
The reaction rate is then prescribed through a forward affinity kinetics law
\begin{align}
    \hat{r}_{\alpha}^{\mathrm{reac},\eta} = k_{\alpha}^{\eta} \bigl\langle \mathcal{B}_{\alpha}^{\eta} \bigr\rangle_{\!+} \,,
\end{align}
with $k_{\alpha}^{\eta} \geq 0$. This choice excludes potentially non-physical reverse reactions. We propose the mobilities $\ell_{\!\alpha \gamma}^{\beta}$ and $k_{\alpha}^{\eta}$ to have the same dependencies as in \eqref{eq:class_concentration_exchange} and to be degenerate with vanishing reactants. In addition, the host-switching mobility $\ell_{\!\alpha \gamma}^{\beta}$ vanishes whenever either host volume fraction $\phi_{\alpha}$ or $\phi_{\gamma}$ vanishes, since a solute cannot be exchanged with an absent host.

\paragraph{Molar diffusive fluxes.}
For the molar diffusive fluxes, we adopt the constitutive relation
\begin{align}
    \bq_{\alpha}^{\beta} = - \sum_{\gamma=1}^{N} \sum_{\delta=1}^{M} \bK_{\alpha\gamma}^{\beta\delta} \nablax \tilde{\kappa}_{\gamma}^{\delta} \,,
\end{align}
where the mobility tensors are functionals of the same state variables as in \eqref{eq:class_concentration_flux} and satisfy the Onsager reciprocal symmetry. In addition, the block mobility operator associated with the compound indices $(\alpha,\beta)$ is required to be positive semi-definite. More precisely, for every collection of vectors $\boldsymbol{z}_{\alpha}^{\beta} \in \mathbb{R}^{d}$, $\alpha = 1,\dots,N$ and $\beta = 1,\dots,M$, we require
\begin{align}
    \sumalpha \sum_{\gamma=1}^{N} \sumbeta \sum_{\delta=1}^{M} \bigl( \boldsymbol{z}_{\alpha}^{\beta} \bigr)^{\T} \bK_{\alpha\gamma}^{\beta\delta} \boldsymbol{z}_{\gamma}^{\delta} \geq 0 \,.
\end{align}
The mobility tensors are required to be degenerate with respect to the host volume fractions, that is, $\bK_{\alpha\gamma}^{\beta\delta}$ vanishes whenever $\phi_{\alpha}$ or $\phi_{\gamma}$ vanishes, since a solute cannot diffuse within an absent host. Degeneracy with respect to the solute concentration itself is optional. It becomes necessary only if the solute chemical potential is singular as $c_{\alpha}^{\beta} \to 0$. Accordingly, the solute fluxes are driven by gradients of the weighted chemical potentials, and the positivity of the mobility operator guarantees admissible dissipation.

\paragraph{Growth rate.}
For the growth rate tensor, we adopt the linear constitutive relation
\begin{align}
    \bLg = \mathbb{G} : \left( \bFe^{\T} \frac{\partial \Psi}{\partial \bFe} - \sumalpha \varphi_{\alpha} \left( \mu_{\lambda, \alpha} - \sumbeta \tilde{\kappa}_{\alpha}^{\beta} c_{\alpha}^{\beta} \right) \bI \right) \,,
\end{align}
where $\mathbb{G}$ denotes a positive semi-definite fourth-order mobility tensor. The mobility may have the same dependencies as in \eqref{eq:class_growth_rate} and must be degenerate in the sense that it vanishes together with the growing constituents (which are typically one or more of the solid constituents) and solutes required for mass production. This choice identifies the growth rate with the thermodynamic driving tensor appearing in \eqref{eq:constraint_growth_rate}, and therefore guarantees that the growth contribution is dissipative. In combination with the kinematic relation \eqref{eq:growth_rate}, the residual form of the evolution equation for the growth deformation gradient becomes
\begin{align}
    \dot{\bF}_{\!\!\mathrm{g}} - \left( \mathbb{G} : \left( \bFe^{\T} \frac{\partial \Psi}{\partial \bFe} - \sumalpha \varphi_{\alpha} \left( \mu_{\lambda, \alpha} - \sumbeta \tilde{\kappa}_{\alpha}^{\beta} c_{\alpha}^{\beta} \right) \bI \right) \right) \bFg = \boldsymbol{0} \,.
\end{align}

We conclude this section by summarizing all thermodynamic driving forces, fluxes, and resulting constitutive choices in Table~\ref{tab:thermodynamic_forces_fluxes}. It remains to specify the closure relations for the Helmholtz free-energy density and the mobilities
\[
    \Psi \,,\quad
    \mathbb{G} \,,\quad
    \bM_{\alpha \gamma} \,,\quad
    \bK_{\alpha \gamma}^{\beta \delta} \,,\quad
    m_{\xi} \,,\quad
    \ell_{\!\alpha \gamma}^{\beta} \,,\quad
    k_{\alpha}^{\eta} \,.
\]
These must be chosen carefully to represent the physics, biology, or chemistry of the process being modeled.

\begin{table}[pos=t]
    \centering
    \caption{Summary of the thermodynamic driving forces, fluxes or rates, and admissible constitutive choices for the individual dissipative processes.}
    \renewcommand{\arraystretch}{1.25}
    \begin{tabular}{llll}
        \toprule
        Process & Thermodynamic force/affinity & Flux/rate & Constitutive choice \\
        \midrule
        phenotype transition & $\mathcal A_{\xi}$ & $\hat{\zeta}^{\exchange}_{\xi}$ & forward affinity kinetics \\
        \addlinespace
        constituent transport & $\rho_{\alpha}^{-1} \bigl( \nablax \mu_{\lambda,\alpha} - \sumbeta \tilde{\kappa}_{\alpha}^{\beta} \nablax c_{\alpha}^{\beta} \bigr)$ & $\bj_{\!\alpha}$ & cross-coupled Onsager law \\
        \addlinespace
        solute host-switching & $\tilde{\kappa}_{\gamma}^{\beta} - \tilde{\kappa}_{\alpha}^{\beta}$ & $\hat{r}^{\mathrm{sw},\beta}_{\alpha \gamma}$ & symmetric linear exchange \\
        \addlinespace
        local solute reaction & $\mathcal{B}_{\alpha}^{\eta}$ & $\hat{r}_{\alpha}^{\mathrm{reac},\eta}$ & forward affinity kinetics \\
        \addlinespace
        solute diffusion
        & $\nablax \tilde{\kappa}_{\alpha}^{\beta}$ & $\bq_{\alpha}^{\beta}$ & cross-coupled Onsager law \\
        \addlinespace
        volume growth &
        $ \bFe^{\T} \frac{\partial \Psi}{\partial \bFe} - \sumalpha \varphi_{\alpha} \bigl( \mu_{\lambda,\alpha} - \sumbeta \tilde{\kappa}_{\alpha}^{\beta} c_{\alpha}^{\beta} \bigr) \bI$ & $\bLg$ & Onsager-type growth law \\
        \bottomrule
    \end{tabular}
    \label{tab:thermodynamic_forces_fluxes}
\end{table}

\begin{rmk}[Compatibility of the constitutive laws]
Compatibility of the proposed constitutive laws with the restrictions \eqref{eq:constitutive_constraints}, and hence with the dissipation inequality \eqref{eq:dissipation_inequality}, may be verified explicitly by inserting the constitutive models into the separate process inequalities.
This yields
\begin{subequations}
    \begin{align}
        \sum_{\xi \in \mathcal{T}} m_{\xi} \langle \mathcal{A}_{\xi} \rangle_{\!+}^{2} &\geq 0 \,,\\
        \sumalpha \sum_{\gamma = 1}^{N} \rho_{\alpha}^{-1} \rho_{\gamma}^{-1} \left( \nablax \mu_{\lambda, \alpha} - \sumbeta \tilde{\kappa}_{\alpha}^{\beta} \nablax c_{\alpha}^{\beta} \right)^{\T} \bM_{\alpha\gamma} \left( \nablax \mu_{\lambda, \gamma} - \sum_{\beta=1}^{M} \tilde{\kappa}_{\gamma}^{\beta} \nablax c_{\gamma}^{\beta} \right) &\geq 0 \,,\\
        \sumalpha \sum_{\gamma = 1}^{N} \sumbeta \frac{1}{2} \ell_{\!\alpha\gamma}^{\beta} \bigl( \tilde{\kappa}_{\gamma}^{\beta} - \tilde{\kappa}_{\alpha}^{\beta} \bigr)^{2}  + \sumalpha \sum_{\eta \in \mathcal{Z}_{\alpha}} k_{\alpha}^{\eta} \bigl\langle \mathcal{B}_{\alpha}^{\eta} \bigr\rangle_{\!+}^{2} &\geq 0\label{eq:dissipation_verification_reactions}  \,,\\
        \sumalpha \sum_{\gamma = 1}^{N} \sumbeta \sum_{\delta = 1}^{M} \bigl( \nablax \tilde{\kappa}_{\alpha}^{\beta} \bigr)^{\T} \bK_{\alpha \gamma}^{\beta \delta} \nablax \tilde{\kappa}_{\gamma}^{\delta} &\geq 0 \,,\\
        \left( \sumalpha \varphi_{\alpha} \left( \mu_{\lambda, \alpha} - \sumbeta \tilde{\kappa}_{\alpha}^{\beta} c_{\alpha}^{\beta} \right) \bI  - \bFe^{\T} \frac{\partial \Psi}{\partial \bFe} \right) : \mathbb{G} : \left( \sumalpha \varphi_{\alpha} \left( \mu_{\lambda, \alpha} - \sumbeta \tilde{\kappa}_{\alpha}^{\beta} c_{\alpha}^{\beta} \right) \bI  - \bFe^{\T} \frac{\partial \Psi}{\partial \bFe} \right) &\geq 0 \,.
    \end{align}
\end{subequations}
\end{rmk}

\section{Numerical implementation}\label{sec:implementation}
In this section, we briefly detail the numerical implementation of the full multiphase growth model derived above. Because growth continuously alters the current configuration, a direct spatial discretization would have to track a moving, growing domain. We therefore recast the model on the fixed reference configuration of the solid skeleton before deriving its weak form. In Section~\ref{sec:total_lagrangian_formulation}, we first reformulate the governing equations in a total Lagrangian setting. Subsequently, in Section~\ref{sec:weak_formulation}, we derive the weak form of the resulting mixed problem, providing the basis for a finite element implementation in solvers such as \emph{FEniCSx} \citep{Fenicsx_3, Fenicsx_2, Fenicsx_1}.

\subsection{Total Lagrangian formulation}\label{sec:total_lagrangian_formulation}
For the numerical treatment of the governing equations, we pull the system back to the reference configuration $\Oreference$ of the solid skeleton. Since the mixture deformation is described by the common solid-skeleton map, all balance equations are expressed with respect to the material points $\bX \in \Oreference$. We first introduce the Helmholtz free-energy density per unit reference volume as a functional of the constitutive class
\begin{align}
    \Psi_{0} = \Psi_{0} \Bigl( \bFe, \bFg, \bigl\{ c_{\alpha}^{\beta} \bigr\}, \bigl\{ \phi_{\alpha} \bigr\}, \bigl\{ \nablaX \phi_{\alpha} \bigr\} \Bigr) \,,
\end{align}
such that
\begin{align}
    \Psi_{0} = \Js \Psi \Bigl( \bFe, \bigl\{ c_{\alpha}^{\beta} \bigr\}, \bigl\{ \phi_{\alpha} \bigr\}, \bigl\{ \nablax \phi_{\alpha} \bigr\} \Bigr) \,,
\end{align}
where the spatial and material gradient operators are connected through the Piola transformation $\nablax \phi_{\alpha} = \bFs^{-\T} \nablaX \phi_{\alpha}$. Accordingly, the Lagrangian chemical potentials are defined by
\begin{subequations}
    \begin{align}
        \mu_{0, \alpha} &\coloneq \frac{\partial \Psi_{0}}{\partial \phi_{\alpha}} - \Div \left( \frac{\partial \Psi_{0}}{\partial \nablaX \phi_{\alpha}} \right)\label{eq:reference_chemical_potential} \,,\\
        \mu_{0\lambda,\alpha} &\coloneq \mu_{0,\alpha} + \Js \lambda \,,\\
        \tilde{\kappa}_{0,\alpha}^{\beta} &\coloneq \phi_{\alpha}^{-1} \frac{\partial \Psi_{0}}{\partial c_{\alpha}^{\beta}} \,,
    \end{align}
\end{subequations}
with $\Div ( \bullet )$ denoting the divergence operator with respect to the material coordinates. The first Piola--Kirchhoff stress tensor is obtained from the standard Piola transformation
\begin{align}
    \bP \coloneq \Js \bsigma \bFs^{-\T} \,.
\end{align}
Using the constitutive expression for the Cauchy stress \eqref{eq:cauchy_stress}, this can equivalently be written as
\begin{align}
    \bP = \frac{\partial \Psi_{0}}{\partial \bFe} \bFg^{-\T} - \sumalpha \phi_{\alpha} \mu_{0\lambda, \alpha} \bFs^{-\T} \,.
    \label{eq:PK1_stress}
\end{align}
Moreover, to express the transport equations on $\Oreference$, we introduce the Piola transforms of the constituent mass fluxes and solute fluxes,
\begin{align}
    \bJ_{\!\alpha} \coloneq \Js \bFs^{-1} \bj_{\!\alpha} \,, \qquad \bQ_{\alpha}^{\beta} \coloneq \Js \bFs^{-1} \bq_{\alpha}^{\beta} \,,
\end{align}
as well as the pulled-back source terms
\begin{align}
    Z_{\alpha}^{\exchange} \coloneq \Js \zeta_{\alpha}^{\exchange} \,, \qquad R_{\alpha}^{\beta} \coloneq \Js r_{\alpha}^{\beta} \,. 
\end{align}
With these definitions, the Lagrangian forms of the constitutive flux laws become
\begin{subequations}
    \begin{align}
        \bJ_{\!\alpha} &= - \sum_{\gamma=1}^{N} \bM_{0, \alpha\gamma}  \rho_{\gamma}^{-1} \left( \nablaX \bigl( \Js^{-1} \mu_{0\lambda, \gamma} \bigr) - \sum_{\beta=1}^{M} \Js^{-1} \tilde{\kappa}_{0,\gamma}^{\beta} \nablaX c_{\gamma}^{\beta} \right) \,,\\
        \bQ_{\alpha}^{\beta} &= - \sum_{\gamma=1}^{N} \sum_{\delta=1}^{M} \bK_{0,\alpha\gamma}^{\beta\delta} \nablaX \bigl( \Js^{-1} \tilde{\kappa}_{0,\gamma}^{\delta} \bigr) \,,
    \end{align}
\end{subequations}
where the pulled-back mobility tensors are given by
\begin{align}
    \bM_{0, \alpha\gamma} \coloneq \Js \bFs^{-1} \bM_{\alpha\gamma} \bFs^{-\T} \,,\qquad \bK_{0,\alpha\gamma}^{\beta\delta} \coloneq \Js \bFs^{-1} \bK_{\alpha\gamma}^{\beta\delta} \bFs^{-\T} \,.
\end{align}
The exchange and reaction terms can likewise be expressed in terms of Lagrangian energetic quantities as
\begin{subequations}
    \begin{align}
        Z_{\alpha}^{\exchange} &= \sum_{\xi \in \mathcal{T}} s_{\alpha,\xi} m_{0,\xi} \bigl\langle \mathcal{A}_{0,\xi} \bigr\rangle_{\!+}\label{eq:constitutive_choice_mass_exchange} \,,\\
        R_{\alpha}^{\beta} &= \sum_{\gamma = 1}^{N} \ell_{0,\alpha\gamma}^{\beta} \Js^{-1} \bigl( \tilde{\kappa}_{0,\gamma}^{\beta} - \tilde{\kappa}_{0,\alpha}^{\beta} \bigr) + \sum_{\eta \in \mathcal{Z}_{\alpha}} \nu_{\alpha}^{\beta,\eta} k_{0,\alpha}^{\eta} \bigl\langle \mathcal{B}_{0,\alpha}^{\eta} \bigr\rangle_{\!+} \,,
    \end{align}
\end{subequations}
with the Lagrangian affinities
\begin{align}
    \mathcal{A}_{0,\xi} \coloneq - \sumalpha \rho_{\alpha}^{-1} s_{\alpha,\xi} \left( \Js^{-1} \mu_{0\lambda, \alpha} - \sumbeta \Js^{-1} \tilde{\kappa}_{0,\alpha}^{\beta} c_{\alpha}^{\beta} \right) \,,\qquad \mathcal{B}_{0,\alpha}^{\eta} \coloneq - \sum_{\beta=1}^{M} \nu_{\alpha}^{\beta,\eta} \Js^{-1} \tilde{\kappa}_{0,\alpha}^{\beta}\label{eq:reference_affinities} \,,
\end{align}
and mobilities
\begin{align}
    m_{0,\xi} \coloneq \Js m_{\xi} \,,\qquad \ell_{0,\alpha\gamma}^{\beta} \coloneq \Js \ell_{\alpha\gamma}^{\beta} \,,\qquad k_{0,\alpha}^{\eta} \coloneq \Js k_{\alpha}^{\eta} \,.
\end{align}
The growth rate tensor can also be formulated in terms of Lagrangian energetic quantities using the transformation
\begin{align}
    \Js \bFe^{\T} \frac{\partial \Psi}{\partial \bFe} = \bFe^{\T} \frac{\partial \Psi_{0}}{\partial \bFe} - \Psi_{0} \bI + \sumalpha \bigl( \bFg^{-\T} \nablaX \phi_{\alpha} \bigr) \otimes \left( \bFg \frac{\partial \Psi_{0}}{\partial \nablaX \phi_{\alpha}} \right) \,.
\end{align}
We additionally introduce the Eshelby stress tensor
\begin{align}
    \bSigma \coloneq \Psi_{0} \bI - \bFe^{\T} \frac{\partial \Psi_{0}}{\partial \bFe} \,,
\end{align}
such that the growth law becomes
\begin{align}
    \bLg = - \mathbb{G}_{0} : \left( \bSigma - \sumalpha \bigl( \bFg^{-\T} \nablaX \phi_{\alpha} \bigr) \otimes \left( \bFg \frac{\partial \Psi_{0}}{\partial \nablaX \phi_{\alpha}} \right) + \sumalpha \varphi_{\alpha} \left( \mu_{0\lambda, \alpha} - \sumbeta \tilde{\kappa}_{0,\alpha}^{\beta} c_{\alpha}^{\beta} \right) \bI \right)\label{eq:reference_growth_law} \,,
\end{align}
where $\mathbb{G}_{0} \coloneq \Js^{-1} \mathbb{G}$ denotes the corresponding Lagrangian growth mobility. The appearance of the Eshelby stress highlights the configurational nature of the thermodynamic driving force for growth and is consistent with related formulations in growth mechanics, see, e.g., \cite{DiCarlo_2002, Ambrosi_2007, Xue_2016}.

The Jacobian of the deformation gradient satisfies $\dot{J}_{\!\mathrm{s}} = \Js \div ( \bvs )$, which allows us to transform the Eulerian system of balance laws \eqref{eq:eulerian_problem_formulation} into its total Lagrangian counterpart.
We use the standard Piola transformations to express spatial differential operators with respect to the reference configuration. The governing equations for all material points $\bX \in \Oreference$ and time $t \geq 0$ take the form
\begin{subequations}
    \begin{align}
        \dt \bigl( \Js \tilde{\rho}_{\alpha} \bigr) + \Div ( \bJ_{\!\alpha} ) - Z_{\alpha}^{\exchange} - \Js \varphi_{\alpha} \rho_{\alpha} \tr ( \bLg ) &= 0 \,, \\
        \dt \bigl( \Js \tilde{c}_{\alpha}^{\beta} \bigr) + \Div \bigl( \tilde{\rho}_{\alpha}^{-1} \tilde{c}_{\alpha}^{\beta} \bJ_{\!\alpha} + \bQ_{\alpha}^{\beta} \bigr) - R_{\alpha}^{\beta} &= 0 \,,\\
        \Div ( \bP ) &= \boldsymbol{0} \,.
    \end{align}
    \label{eq:balance_laws_reference}
\end{subequations}
The saturation constraint is enforced through its reference rate form. Starting from \eqref{eq:saturation_constraint_rate_form_2} and using the pulled-back quantities defined above gives
\begin{align}
    \frac{\d}{\d t} \Js + \sumalpha \rho_{\alpha}^{-1} \bigl( \Div ( \bJ_{\!\alpha} ) - Z_{\alpha}^{\exchange} - \Js \varphi_{\alpha} \rho_{\alpha} \tr ( \bLg ) \bigr) = 0 \,.
    \label{eq:saturation_constraint_reference}
\end{align}
Together with the initial condition $\sumalpha \phi_{\alpha} ( \bX, t=0 ) = 1$, this ensures that full saturation is preserved for all $t \geq 0$.

\subsection{Weak formulation}\label{sec:weak_formulation}
We now derive the weak form of the total Lagrangian problem stated in Section~\ref{sec:total_lagrangian_formulation}. The formulation is written as a fully coupled mixed problem for the volume fractions, solute concentrations, constituent chemical potentials, saturation multiplier, displacement field, and growth deformation gradient. The chemical potentials are introduced as separate unknown fields, thereby reducing the fourth-order mass balance equations to a coupled system of second-order equations. We treat the growth deformation gradient not as a local internal variable, but as an additional unknown tensor field. Although this increases the size of the global nonlinear system, it simplifies the implementation of the model in existing finite element frameworks, where all unknowns can be handled as components of a mixed function. Since computational efficiency is not the primary objective of this work, we adopt this formulation for convenience.

We collect the unknown fields and corresponding test functions in the tuples
\begin{align}
    U ( t ) \coloneq \Bigl( \bigl\{ \phi_{\alpha} \bigr\} , \bigl\{ c_{\alpha}^{\beta} \bigr\} , \bigl\{ \mu_{0,\alpha} \bigr\} , \lambda \,, \bu \,, \bFg \Bigr) \,,\qquad \delta U \coloneq \Bigl( \bigl\{ \delta \phi_{\alpha} \bigr\} , \bigl\{ \delta c_{\alpha}^{\beta} \bigr\} , \bigl\{ \delta \mu_{0,\alpha} \bigr\} , \delta \lambda \,, \delta \bu \,, \delta \bFg \Bigr) \,,
\end{align}
respectively. The first four unknowns are assumed to belong to admissible spaces with the regularity required by the weak form. In particular, we take
\begin{align}
    \mathcal{V}_{\phi} \subset H^{1} ( \Oreference ) \,,\quad \mathcal{V}_{c} \subset H^{1} ( \Oreference ) \,,\quad \mathcal{V}_{\mu} \subset H^{1} ( \Oreference ) \,,\quad \mathcal{V}_{\bu} \subset [ H^{1} ( \Oreference ) ]^{d} \,,\quad \mathcal{V}_{\lambda} \subset H^{1} ( \Oreference ) \,,
\end{align}
where $H^{1} ( \Oreference )$ denotes the usual Sobolev space of square-integrable functions with square-integrable weak gradients. Since the growth evolution equation does not involve spatial gradients of the growth deformation gradient, only square-integrability is required for this field. We therefore additionally introduce the tensor-valued space
\begin{align}
    \mathcal{V}_{\mathrm{g}} \subset [ L^{2} ( \Oreference ) ]^{d \times d} \,.
\end{align}
The corresponding mixed trial space is constructed as
\begin{align}
    \mathcal{V} \coloneq \left( \prod_{\alpha = 1}^{N} \mathcal{V}_{\phi} \right) \times \left( \prod_{\alpha = 1}^{N} \prod_{\beta = 1}^{M} \mathcal{V}_{c} \right) \times  \left( \prod_{\alpha = 1}^{N} \mathcal{V}_{\mu} \right) \times \mathcal{V}_{\lambda} \times \mathcal{V}_{\bu} \times \mathcal{V}_{\mathrm{g}} \,.
\end{align}

The boundary of the reference domain is denoted by $\partial \Oreference$, with outward unit normal vector $\bN$ and surface element $\d A_{0}$, while volume elements of $\Oreference$ are denoted by $\d V_{0}$. Multiplying the governing equations \eqref{eq:balance_laws_reference}, the constituent chemical potentials \eqref{eq:reference_chemical_potential}, the saturation constraint \eqref{eq:saturation_constraint_reference}, and the growth law residual $\dot{\bF}_{\!\!\mathrm{g}} - \bLg \bFg = \boldsymbol{0}$ by the corresponding test functions, integrating over the reference domain, and applying integration by parts to the divergence terms yields the residual equations
\begin{subequations}
    \begin{align}
        \begin{split}
            \mathcal{R}_{\phi} &\coloneq \sumalpha \int_{\Oreference} \left( \dt \bigl( \Js \tilde{\rho}_{\alpha} \bigr) - Z_{\alpha}^{\exchange} - \Js \varphi_{\alpha} \rho_{\alpha} \tr ( \bLg ) \right) \delta \phi_{\alpha} - \bJ_{\!\alpha} \cdot \nablaX ( \delta \phi_{\alpha} ) \,\d V_{0} \\
            &\qquad+ \sumalpha \int_{\partial \Oreference} \bJ_{\!\alpha} \cdot \bN \delta \phi_{\alpha} \,\d A_{0} \,,
        \end{split} \\
        \begin{split}
        \mathcal{R}_{c} &\coloneq \sumalpha \sumbeta \int_{\Oreference} \left( \dt \bigl( \Js \tilde{c}_{\alpha}^{\beta} \bigr) - R_{\alpha}^{\beta} \right) \delta c_{\alpha}^{\beta} - \bigl( \tilde{\rho}_{\alpha}^{-1} \tilde{c}_{\alpha}^{\beta} \bJ_{\!\alpha} + \bQ_{\alpha}^{\beta} \bigr) \cdot \nablaX \bigl( \delta c_{\alpha}^{\beta} \bigr) \,\d V_{0} \\
        &\qquad+ \sumalpha \sumbeta \int_{\partial \Oreference} \bigl( \tilde{\rho}_{\alpha}^{-1} \tilde{c}_{\alpha}^{\beta} \bJ_{\!\alpha} + \bQ_{\alpha}^{\beta} \bigr) \cdot \bN \delta c_{\alpha}^{\beta} \,\d A_{0} \,,
        \end{split} \\
        \mathcal{R}_{\mu} &\coloneq \sumalpha \int_{\Oreference} \left( \mu_{0,\alpha} - \frac{\partial \Psi_{0}}{\partial \phi_{\alpha}} \right) \delta \mu_{0,\alpha} - \frac{\partial \Psi_{0}}{\partial \nablaX \phi_{\alpha}} \cdot \nablaX ( \delta \mu_{0,\alpha} ) \,\d V_{0} + \sumalpha \int_{\partial \Oreference} \frac{\partial \Psi_{0}}{\partial \nablaX \phi_{\alpha}} \cdot \bN \delta \mu_{0,\alpha} \,\d A_{0} \,, \\
        \begin{split}
            \mathcal{R}_{\lambda} &\coloneq \int_{\Oreference} \left( \frac{\d}{\d t} \Js - \sumalpha \left( \rho_{\alpha}^{-1} Z_{\alpha}^{\exchange} + \Js \varphi_{\alpha} \tr ( \bLg ) \right) \right) \delta \lambda - \sumalpha \rho_{\alpha}^{-1} \bJ_{\!\alpha} \cdot \nablaX ( \delta \lambda ) \,\d V_{0} \\
            &\qquad+ \int_{\partial \Oreference} \sumalpha \rho_{\alpha}^{-1} \bJ_{\!\alpha} \cdot \bN \delta \lambda \,\d A_{0} \,, 
        \end{split} \\
        \mathcal{R}_{\bu} &\coloneq \int_{\Oreference} \bP : \nablaX ( \delta \bu ) \,\d V_{0} - \int_{\partial \Oreference} ( \bP \bN ) \cdot \delta \bu \,\d A_{0} \,, \\
        \mathcal{R}_{\mathrm{g}} &\coloneq \int_{\Oreference} \left( \frac{\d}{\d t} \bFg - \bLg \bFg \right) : \delta \bFg \,\d V_{0} \,.
    \end{align}
\end{subequations}
Here, $\mathcal{R}_{\phi}$ represents the weak form of the constituent mass balances, $\mathcal{R}_{c}$ corresponds to the solute concentration balances, $\mathcal{R}_{\lambda}$ enforces the rate form of the saturation constraint, $\mathcal{R}_{\bu}$ is the mechanical equilibrium residual, and $\mathcal{R}_{\mathrm{g}}$ is the residual of the growth evolution equation. The total residual of the coupled mixed problem is then defined as
\begin{align}
    \mathcal{R} ( U ( t ), \delta U ) \coloneq \mathcal{R}_{\phi} + \mathcal{R}_{c} + \mathcal{R}_{\mu} + \mathcal{R}_{\bu} + \mathcal{R}_{\lambda} + \mathcal{R}_{\mathrm{g}} \,.
\end{align}

The weak form of the problem is stated as follows: for all $t \in ( 0, T ]$, find $U ( t ) \in \mathcal{V}$ such that $\mathcal{R} ( U ( t ), \delta U ) = 0$ for all $\delta U \in \mathcal{V}_{0}$. Here, $T$ is the maximum time and $\mathcal{V}_{0}$ denotes the test space associated with the mixed trial space $\mathcal{V}$. The subscript $0$ indicates that the test functions satisfy homogeneous essential boundary conditions on those parts of the boundary where the corresponding trial fields are prescribed.

The weak problem is supplemented by suitable initial data $U ( t = 0 ) = \bar{U}$. Essential boundary conditions are incorporated in the definition of the trial and test spaces, whereas natural boundary conditions enter through the boundary terms appearing in the residuals. For the temporal discretization, the time derivatives appearing in the residuals are approximated by an implicit Euler scheme, which evaluates all nonlinear terms at the new time level and thereby leads to a nonlinear algebraic system at each time step. For details on the spatial and temporal discretization procedure as well as the linearization and solution of nonlinear systems, we refer, for example, to \cite{Hughes_1987, Wriggers_2008}.

\section{Specialization to avascular tumor growth}\label{sec:tumor_model}
\begin{figure}[pos=htbp]
    \centering
    \includegraphics{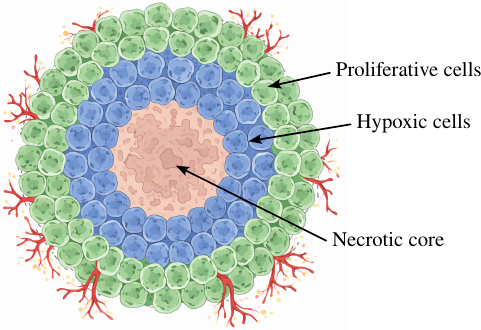}
    \caption{Schematic of an avascular tumor spheroid. Oxygen and other nutrients supplied by the surrounding host vasculature diffuse inward from the tumor boundary and are progressively consumed, so that their concentration decreases toward the center. The resulting gradient organizes the tumor into three concentric regions: a well-oxygenated, proliferating outer rim that sustains volumetric growth, an intermediate hypoxic zone of oxygen-deprived cells, and a necrotic core where prolonged oxygen depletion has led to cell death. The illustration was generated with \emph{GPT-5.6 Sol} and edited by the authors.}
    \label{fig:tumor_spheroid}%
\end{figure}

We now demonstrate the general framework by specializing it to a concrete model of avascular tumor growth. In the absence of a vascular supply, nutrients reach the tumor only by diffusion from the surrounding tissue, so that a proliferating outer rim, a hypoxic intermediate zone, and a necrotic core emerge as the nutrient level decreases toward the interior \citep{Folkman_1973, Sutherland_1988}. The formation of this characteristic pattern is illustrated in Figure~\ref{fig:tumor_spheroid}. The guiding idea of our model is that, once the constituents, the exchange and reaction channels, the free-energy density, and the mobilities are fixed, all remaining constitutive quantities, namely chemical potentials, stresses, affinities, fluxes, and the growth law, follow by insertion into the general relations derived above. Here, we therefore specify the model-defining ingredients only, while the explicit evaluations of chemical potentials, stresses, and related quantities are collected in Appendix~\ref{sec:tumor_constitutive}. To the
best of our knowledge, this is the first avascular tumor model in which the phenotype transitions between proliferative, hypoxic, and necrotic cells and the consumption of nutrients are governed by thermodynamically consistent affinity kinetics. These processes are embedded in finite growth kinematics, so that phenotype changes, volume growth, and growth-induced residual stress follow from one common free-energy dissipation structure.

We consider a fully saturated mixture of $N = 4$ constituents. The solid skeleton is formed by tumor cells that are distinguished by their phenotype into proliferative ($\mathrm{p}$), hypoxic ($\mathrm{h}$), and necrotic ($\mathrm{n}$) cells, while the pore space is filled by an interstitial fluid ($\mathrm{f}$),
\begin{align}
    \alpha \in \{ \mathrm{p}, \mathrm{h}, \mathrm{n}, \mathrm{f} \} \,, \qquad \phi_{\mathrm{s}} \coloneq \phip + \phih + \phin \,, \qquad \phip + \phih + \phin + \phif = 1 \,,
\end{align}
where $\phi_{\mathrm{s}}$ denotes the total solid volume fraction and the last relation is the saturation constraint \eqref{eq:saturation_constraint}. The three cellular phenotypes share the same intrinsic mass density $\rho_{\mathrm{s}}$, whereas the fluid has intrinsic density $\rho_{\mathrm{f}}$. The fluid hosts $M = 2$ dilute solutes, namely oxygen ($\mathrm{o}$) acting as a nutrient and a generic metabolic waste product ($\mathrm{w}$),
\begin{align}
    \beta \in \{ \mathrm{o}, \mathrm{w} \} \,, \qquad
    c_{\mathrm{p}}^{\beta} = c_{\mathrm{h}}^{\beta} = c_{\mathrm{n}}^{\beta} = 0\label{eq:tumor_solutes_choice} \,,
\end{align}
so that the only non-vanishing molar concentrations are $\co$ and $\cw$.

\begin{figure}[pos=htbp]
    \centering
    \includegraphics{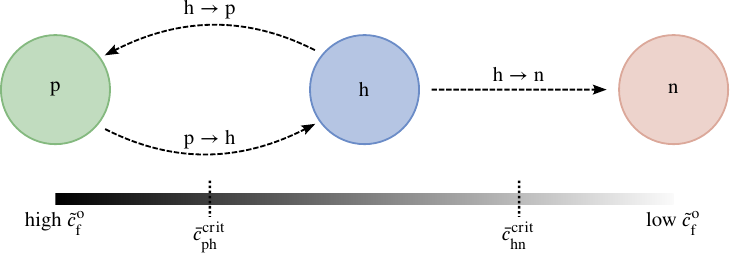}
    \caption{Phenotype transition network of the avascular tumor model, governed by the local oxygen concentration $\tilde{c}_{\mathrm{f}}^{\mathrm{o}}$. Tumor cells switch between proliferative ($\mathrm{p}$), hypoxic ($\mathrm{h}$), and necrotic ($\mathrm{n}$) state. Proliferative cells become hypoxic once the oxygen level falls below a threshold $\bar{c}_{\mathrm{ph}}^{\mathrm{crit}}$ and recover upon re-oxygenation. The hypoxic--necrotic transition is irreversible and is triggered once the oxygen level drops below $\bar{c}_{\mathrm{hn}}^{\mathrm{crit}}$.}
    \label{fig:transitions}%
\end{figure}

\paragraph{Transition and reaction channels.}
Cell phenotype changes are modeled as inter-constituent mass exchange between cells. We admit the transition channels
\begin{align}
    \mathcal{T} \coloneq \{ \mathrm{p} \to \mathrm{h} \,,\, \mathrm{h} \to \mathrm{p} \,,\, \mathrm{h} \to \mathrm{n} \} \,,
\end{align}
that is, proliferative cells may become hypoxic under nutrient deprivation and recover under re-oxygenation, while hypoxic cells may irreversibly become necrotic. Each channel $\xi \in \mathcal{T}$ is characterized by its stoichiometric exchange vector $\boldsymbol{s}_{\xi} \coloneq ( s_{\mathrm{p},\xi}, s_{\mathrm{h},\xi}, s_{\mathrm{n},\xi}, s_{\mathrm{f},\xi} )^{\T}$, with a $-1$ entry for the consumed and a $+1$ entry for the produced phenotype,
\begin{align}
    \boldsymbol{s}_{\mathrm{p} \to \mathrm{h}} \coloneq ( -1, +1, 0, 0 )^{\T} \,, \qquad
    \boldsymbol{s}_{\mathrm{h} \to \mathrm{p}} \coloneq ( +1, -1, 0, 0 )^{\T} \,, \qquad
    \boldsymbol{s}_{\mathrm{h} \to \mathrm{n}} \coloneq ( 0, -1, +1, 0 )^{\T} \,,
\end{align}
which satisfy the mass-conservation condition $\sum_{\alpha} s_{\alpha,\xi} = 0$. This transition network is illustrated in Figure~\ref{fig:transitions}. Within the fluid, oxygen is consumed and converted into waste through the single reaction channel
\begin{align}
    \mathcal{Z}_{\mathrm{f}} \coloneq \{ \mathrm{o} \to \mathrm{w} \} \,, \qquad
    \boldsymbol{\nu}_{\mathrm{f}}^{\mathrm{o} \to \mathrm{w}} \coloneq \left( -1, \tfrac{\mathsf{M}^{\mathrm{o}}}{\mathsf{M}^{\mathrm{w}}} \right)^{\T} \,,
\end{align}
where the stoichiometric coefficients are chosen to satisfy the mass-conservation constraint $\sum_{\beta} \mathsf{M}^{\beta} \nu_{\mathrm{f}}^{\beta, \mathrm{o} \to \mathrm{w}} = 0$.

\paragraph{Free-energy density.}
The Helmholtz free-energy density per unit reference volume is additively decomposed into a mechanical, a mixing, a fitness, and an interfacial contribution,
\begin{align}
    \Psi_{0} = \Psi_{0}^{\mathrm{mech}} + \Psi_{0}^{\mathrm{mix}} + \Psi_{0}^{\mathrm{fit}} + \Psi_{0}^{\internal} \,.
\end{align}
The mechanical part is a compressible neo-Hookean energy of the elastically deformed solid skeleton,
\begin{align}
    \Psi_{0}^{\mathrm{mech}} \coloneq \frac{G}{2} ( \tr ( \bCe ) - d - 2 \log ( \Je ) ) \,,
    \label{eq:tumor_psi_mech}
\end{align}
where $G > 0$ is the shear modulus and $\bCe \coloneq \bFe^{\T} \bFe$ is the elastic right Cauchy--Green tensor. This contribution generates the elastic and growth-induced residual stresses. We assume adhesive properties between tumor cells such that their mixing energy contribution is formulated entirely through the solid volume fraction $\phi_{\s}$ \citep{Wise_2008, Lima_2014}. Thus, the mixing part collects the entropic mixing of solid and fluid together with the quadratic energetic penalties of the dissolved solutes,
\begin{align}
    \Psi_{0}^{\mathrm{mix}} \coloneq \Js \left( E_{\mathrm{s}} \phi_{\mathrm{s}} \log ( \phi_{\mathrm{s}} ) + E_{\mathrm{f}} \phif \log ( \phif ) + \frac{A}{c^{*}} \tilde{c}_{\mathrm{f}}^{\mathrm{o}} \phi_{\mathrm{s}} + \frac{\phif \epsilon^{\mathrm{o}}}{2} \left( \frac{\co}{c^{*}} \right)^{2} + \frac{\phif \epsilon^{\mathrm{w}}}{2} \left( \frac{\cw}{c^{*}} \right)^{2} \right) \,,
    \label{eq:tumor_psi_mix}
\end{align}
where $E_{\mathrm{s}}$ and $E_{\mathrm{f}}$ are mixing moduli, $A$ measures the affinity of oxygen for the cellular phase, $c^{*}$ is a spatially uniform reference concentration, and $\epsilon^{\mathrm{o}}, \epsilon^{\mathrm{w}}$ penalize deviations of the solute concentrations. We recall that $\tilde{c}_{\mathrm{f}}^{\mathrm{o}} = \phif \co$ denotes the oxygen concentration per unit mixture volume, and all parameters are positive constants.
\begin{rmk}[Regularization of entropic mixing energy]
    The derivative of the entropic mixing energy $\phi_{\alpha} \log ( \phi_{\alpha} )$ becomes singular as $\phi_{\s} \to 0$ and $\phif \to 0$, which causes numerical instabilities in the finite element implementation of the model. We therefore regularize this term with a suitable polynomial approximation \citep{Eikelder_2026}. Details on this are given in Appendix~\ref{sec:polynomial_approximation}.
\end{rmk}
The fitness part encodes the phenotype response to the local oxygen level through two critical concentrations,
\begin{align}
    \Psi_{0}^{\mathrm{fit}} \coloneq \Js \left( \frac{B}{c^{*}} \bigl( \bar{c}_{\mathrm{ph}}^{\mathrm{crit}} - \tilde{c}_{\mathrm{f}}^{\mathrm{o}} \bigr) \phip + \frac{B}{c^{*}} \bigl( \tilde{c}_{\mathrm{f}}^{\mathrm{o}} - \bar{c}_{\mathrm{hn}}^{\mathrm{crit}} \bigr) \phin \right) \,,
    \label{eq:tumor_psi_fit}
\end{align}
with fitness modulus $B > 0$ and positive thresholds $\bar{c}_{\mathrm{ph}}^{\mathrm{crit}}$ and $\bar{c}_{\mathrm{hn}}^{\mathrm{crit}} < \bar{c}_{\mathrm{ph}}^{\mathrm{crit}}$ governing the proliferative--hypoxic and hypoxic--necrotic transitions, respectively. Finally, the interfacial part penalizes sharp gradients of the cellular phenotype fractions and thereby regularizes the phase boundaries,
\begin{align}
    \Psi_{0}^{\internal} \coloneq \Js \frac{\gamma}{2} \Bigl( \bigl\lVert \bFs^{-\T} \nablaX \phip \bigr\rVert^{2} + \bigl\lVert \bFs^{-\T} \nablaX \phih \bigr\rVert^{2} + \bigl\lVert \bFs^{-\T} \nablaX \phin \bigr\rVert^{2} \Bigr) \,,
    \label{eq:tumor_psi_int}
\end{align}
where $\gamma$ is the interface parameter and $\lVert \bullet \rVert$ denotes the Euclidean norm. Note that this choice is compatible with condition \eqref{eq:gradient_energy_requirement}.

\paragraph{Fluid transport and solute diffusion.}
Because the solid phenotypes move with the skeleton, their peculiar fluxes vanish, $\bJ_{\!\mathrm{p}} = \bJ_{\!\mathrm{h}} = \bJ_{\!\mathrm{n}} = \boldsymbol{0}$, and only the fluid is transported relative to the skeleton. Choosing $\bM_{0,\alpha\gamma} = \boldsymbol{0}$ if $\alpha \neq \gamma$, the corresponding transport mobility reduces to a single isotropic, Darcy-type permeability of the pore space,
\begin{align}
    \bM_{0,\mathrm{ff}} = \Js \bar{M}_{\mathrm{f}} \phif ( 1 - \phif ) \bC_{\!\mathrm{s}}^{-1} \,,
    \label{eq:tumor_transport_mobility}
\end{align}
that degenerates as the pore space closes with permeability coefficient $\bar{M}_{\mathrm{f}} \geq 0$ and $\bC_{\!\mathrm{s}} \coloneq \bFs^{\T} \bFs$. The dissolved solutes diffuse within the fluid according to a Fick-type law obtained from a diagonal, isotropic diffusion mobility,
\begin{align}
    \bK_{0,\mathrm{ff}}^{\mathrm{oo}} = \Js \bar{D}^{\mathrm{o}} \bigl( \tilde{c}_{\mathrm{f}}^{\mathrm{o}} \bigr)^{2} \bC_{\!\mathrm{s}}^{-1} \,, \qquad \bK_{0,\mathrm{ff}}^{\mathrm{ww}} = \Js \bar{D}^{\mathrm{w}} \bigl( \phif c^{*} \bigr)^{2} \bC_{\!\mathrm{s}}^{-1} \,,
    \label{eq:tumor_diffusion_mobility}
\end{align}
with diffusion coefficients $\bar{D}^{\mathrm{o}}, \bar{D}^{\mathrm{w}} \geq 0$ and all off-diagonal solute couplings set to zero. The explicit constituent flux $\bJ_{\!\mathrm{f}}$ and the solute diffusive fluxes $\bQ_{\mathrm{f}}^{\mathrm{o}}$, $\bQ_{\mathrm{f}}^{\mathrm{w}}$, obtained by inserting the potentials into the transport laws of Section~\ref{sec:total_lagrangian_formulation}, are collected in Appendix~\ref{sec:tumor_constitutive}.

\paragraph{Cell phenotype transitions and oxygen consumption.}
Since the solutes reside exclusively in the fluid, the transition affinities \eqref{eq:reference_affinities} involve only the cellular chemical potentials and evaluate to
\begin{subequations}
    \begin{align}
        \mathcal{A}_{0,\mathrm{p} \to \mathrm{h}} &= \frac{B}{\rho_{\mathrm{s}} c^{*}} \bigl( \bar{c}_{\mathrm{ph}}^{\mathrm{crit}} - \tilde{c}_{\mathrm{f}}^{\mathrm{o}} \bigr) - \frac{\gamma}{\Js \rho_{\mathrm{s}}} \Div \bigl( \Js \bC_{\!\mathrm{s}}^{-1} \nablaX ( \phip - \phih ) \bigr) \,, \\
        \mathcal{A}_{0,\mathrm{h} \to \mathrm{p}} &= - \mathcal{A}_{0,\mathrm{p} \to \mathrm{h}} \,, \\
        \mathcal{A}_{0,\mathrm{h} \to \mathrm{n}} &= \frac{B}{\rho_{\mathrm{s}} c^{*}} \bigl( \bar{c}_{\mathrm{hn}}^{\mathrm{crit}} - \tilde{c}_{\mathrm{f}}^{\mathrm{o}} \bigr) - \frac{\gamma}{\Js \rho_{\mathrm{s}}} \Div \bigl( \Js \bC_{\!\mathrm{s}}^{-1} \nablaX ( \phih - \phin ) \bigr) \,.
    \end{align}
    \label{eq:tumor_affinities}%
\end{subequations}
The bulk contributions show the intended biology: the proliferative--hypoxic transition is activated whenever the oxygen level drops below $\bar{c}_{\mathrm{ph}}^{\mathrm{crit}}$, and the hypoxic--necrotic transition once it drops below $\bar{c}_{\mathrm{hn}}^{\mathrm{crit}}$, while re-oxygenation reverses the first transition. The non-local divergence terms regularize the formation of sharp interfaces between tumor cells through phenotype transitions. The transition mobilities are chosen to degenerate with the respective reactant phase,
\begin{align}
    m_{0,\mathrm{p} \to \mathrm{h}} = \Js \bar{m}_{\mathrm{ph}} \phip \,,\qquad
    m_{0,\mathrm{h} \to \mathrm{p}} = \Js \bar{m}_{\mathrm{hp}} \phih \,,\qquad
    m_{0,\mathrm{h} \to \mathrm{n}} = \Js \bar{m}_{\mathrm{hn}} \phih \,,
\end{align}
with constant rate coefficients $\bar{m}_{\mathrm{ph}}, \bar{m}_{\mathrm{hp}}, \bar{m}_{\mathrm{hn}} \geq 0$, while the reaction mobility degenerates with the oxygen-consuming cells and the available oxygen,
\begin{align}
    k_{0,\mathrm{f}}^{\mathrm{o} \to \mathrm{w}} = \Js ( \bar{k}_{\mathrm{p}} \phip + \bar{k}_{\mathrm{h}} \phih ) \tilde{c}_{\mathrm{f}}^{\mathrm{o}} \,,
\end{align}
so that proliferative and hypoxic cells actively consume oxygen, whereas necrotic cells have only a passive influence by altering the affinity (see Appendix~\ref{sec:tumor_constitutive}). Applying the forward affinity kinetics \eqref{eq:constitutive_choice_mass_exchange} with these mobilities yields the nominal cellular production terms $Z_{\alpha}^{\exchange}$ (with $Z_{\mathrm{f}}^{\exchange} = 0$, since the fluid does not participate in phenotype transitions) and the solute sources $R_{\mathrm{f}}^{\mathrm{o}}$, $R_{\mathrm{f}}^{\mathrm{w}}$. Their explicit forms are also given in Appendix~\ref{sec:tumor_constitutive}.

\paragraph{Growth kinematics.}
We assume that only proliferative cells produce new volume through mitosis. Consequently, the volume accumulation fractions reduce to
\begin{align}
    \varphi_{\mathrm{p}} = \theta \in [ 0, 1 ] \,, \qquad \varphi_{\mathrm{h}} = \varphi_{\mathrm{n}} = \varphi_{\mathrm{f}} = 0\label{eq:tumor_accumulation_choice} \,,
\end{align}
so that the growth-induced mass production \eqref{eq:growth_coupling} acts on the proliferative constituent alone, while the fraction $1 - \theta$ of the created volume is initially taken up as pore space and subsequently redistributed by fluid transport and elastic accommodation. For the Lagrangian growth mobility we adopt the isotropic fourth-order tensor
\begin{align}
    \mathbb{G}_{0} = g_{0} \mathbb{I} \,, \qquad g_{0} \coloneq \tilde{c}_{\mathrm{f}}^{\mathrm{o}} \phip \frac{\bar{g}}{\Js} \left\langle 1 - \frac{\Jg}{\Jg^{\mathrm{max}}} \right\rangle_{\!\!+} \,, \qquad \bar{g} \geq 0 \,,
    \label{eq:tumor_growth_mobility}
\end{align}
where $\mathbb{I}$ denotes the fourth-order identity tensor. Growth is limited locally by the maximum growth-induced volume increase $\Jg^{\mathrm{max}}$, which we assume corresponds to a tumor size at which proliferation and apoptosis balance. With \eqref{eq:tumor_solutes_choice}, \eqref{eq:tumor_accumulation_choice}, and the Eshelby stress $\bSigma$ evaluated in Appendix~\ref{sec:tumor_constitutive}, the growth law \eqref{eq:reference_growth_law} then reduces to
\begin{align}
    \bLg = - g_{0} \bigl( \Psi_{0}^{\mathrm{mech}} \bI - G ( \bCe - \bI ) + \Js \theta \mu_{\lambda, \mathrm{p}} \bI \bigr) \,.
    \label{eq:tumor_growth_law}
\end{align}
Since $\bLg$ retains the deviatoric part of its driving force, growth is in general anisotropic.

\section{Computational results}\label{sec:results}
We implement the specialized tumor model in the open-source finite element library \emph{FEniCSx} \citep{Fenicsx_3, Fenicsx_2, Fenicsx_1}, building on the total Lagrangian mixed weak form derived in Section~\ref{sec:weak_formulation}. Rather than hard-coding the explicit chemical potentials and stresses derived in Appendix~\ref{sec:tumor_constitutive}, we evaluate them directly from the free-energy density $\Psi_{0}$ by means of the automatic differentiation capabilities of \emph{FEniCSx}. The coupled nonlinear problem is integrated in time with an implicit Euler scheme and solved monolithically with a Newton--Raphson method at each time step.

The remainder of this section is organized as follows. Section~\ref{sec:setup} specifies the geometry, initial state, loading, and boundary conditions of the benchmark problem. Building on this setup, we then investigate three representative growth scenarios that isolate different aspects of the coupled model: unconfined growth (Section~\ref{sec:unconfined}), confined growth (Section~\ref{sec:confined}), and growth under heterogeneous boundary transport (Section~\ref{sec:transport}).

\subsection{Setup of the initial--boundary value problem}\label{sec:setup}
\begin{figure}[pos=htbp]
    \centering
    \includegraphics{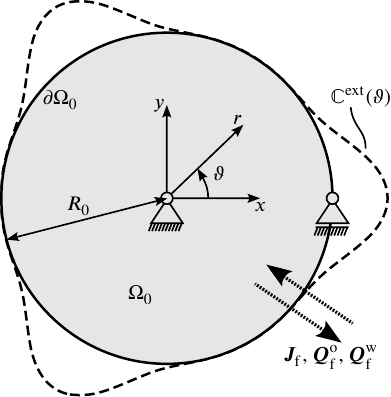}
    \caption{Computational setup for the two-dimensional benchmark problem. Across the boundary, the tumor exchanges fluid and dissolved solutes with the surrounding host reservoir through the permeable-membrane fluxes $\bJ_{\!\mathrm{f}}$, $\bQ_{\mathrm{f}}^{\mathrm{o}}$, and $\bQ_{\mathrm{f}}^{\mathrm{w}}$. In addition, the boundary is elastically supported by the host with an angularly varying stiffness (dashed curve).}
    \label{fig:example_setup}%
\end{figure}

We consider a two-dimensional avascular tumor spheroid with circular reference domain $\Oreference$ of radius $R_{0}$. Material points are addressed either by Cartesian coordinates $(x,y)$ or by polar coordinates $(r,\vartheta)$, as sketched in Figure~\ref{fig:example_setup}. At the initial time ($t = 0$), the non-grown reference domain, $\bFg ( \bX, t = 0 ) = \bI$, is filled by a spatially uniform mixture of proliferative cells and interstitial fluid in equal volume fractions, i.e., $\phip ( \bX, t = 0 ) = 0.5$ and $\phif ( \bX, t = 0 ) = 0.5$. The hypoxic and necrotic phenotypes are initially absent and develop only through the ensuing nutrient dynamics. Furthermore, the tumor is uniformly oxygenated at the reference concentration, $\co ( \bX, t = 0 ) = c^{*}$, and free of metabolic waste. 

\begin{table}[pos=t]
    \centering
    \caption{Numerical values for the non-dimensional parameters used in the tumor growth simulations.}
    \renewcommand{\arraystretch}{1.25}
    \begin{tabular}{llll}
        \toprule
        Symbol & Description & Value \\
        \midrule
        $R_{0}$ & initial tumor radius & $1$ \\
        $G$ & shear modulus of the solid skeleton & $10$ \\
        $E_{\mathrm{s}}$, $E_{\mathrm{f}}$ & mixing moduli of solid and fluid & $100$, $100$ \\
        $A$ & oxygen--cell affinity modulus & $0.02$ \\
        $B$ & phenotype fitness modulus & 0.02 \\
        $\bar{c}_{\mathrm{ph}}^{\mathrm{crit}}$, $\bar{c}_{\mathrm{hn}}^{\mathrm{crit}}$ & critical oxygen concentrations & $0.23$, $0.11$ \\
        $\epsilon^{\mathrm{o}}$, $\epsilon^{\mathrm{w}}$ & solute-energy coefficients & $4$, $0.04$ \\
        $c^{*}$ & reference concentration & $1$ \\
        $\gamma$ & interface parameter & $10^{-4}$ \\
        $\bar{m}_{\mathrm{ph}}$, $\bar{m}_{\mathrm{hp}}$, $\bar{m}_{\mathrm{hn}}$ & phenotype transition rate coefficients & $100$, $100$, $10^{4}$ \\
        $\bar{k}_{\mathrm{p}}$, $\bar{k}_{\mathrm{h}}$ & oxygen consumption rate coefficients & $15.33$, $3.06$ \\
        $\bar{M}_{\mathrm{f}}$ & pore permeability coefficient & 10 \\
        $\bar{D}^{\mathrm{o}}$, $\bar{D}^{\mathrm{w}}$ & solute diffusion coefficients & $4.9$, $100$ \\
        $\bar{g}$ & growth mobility coefficient & $0.02$ \\
        $\Jg^{\mathrm{max}}$ & growth limit of proliferative cells & $9$ \\
        $\theta$ & volume accumulation fraction of proliferative cells & $0.5$ \\
        $\rho_{\mathrm{s}}$, $\rho_{\mathrm{f}}$ & intrinsic densities of cells and fluid & $1$, $1$ \\
        $\mathsf{M}^{\mathrm{o}}$, $\mathsf{M}^{\mathrm{w}}$ & molar masses of oxygen and waste & $1$, $1$ \\
        $\phi_{\s}^{\external}$, $\phi_{\mathrm{f}}^{\external}$ & volume fractions of cells and fluid in the host tissue & $0.5$, $0.5$ \\
        $c_{\mathrm{f}}^{\mathrm{o,ext}}$, $c_{\mathrm{f}}^{\mathrm{w,ext}}$ & oxygen and waste concentrations in the host tissue & $1$, $0$ \\
        $\lambda^{\external}$ & Lagrange multiplier pressure of the host tissue & $-98$ \\
        \bottomrule
    \end{tabular}
    \label{tab:tumor_parameters}
\end{table}

The surrounding host tissue acts simultaneously as a chemical reservoir and as a mechanical support. We model it as a mixture of healthy solid cells with fixed volume fraction $\phi_{\s}^{\external}$ and the same extracellular fluid as in the tumor, with volume fraction $\phi_{\mathrm{f}}^{\external}$. Oxygen and waste are present in the reservoir at prescribed concentrations $c_{\mathrm{f}}^{\mathrm{o,ext}}$ and $c_{\mathrm{f}}^{\mathrm{w,ext}}$, respectively. Oxygen therefore diffuses inward from the reservoir and is consumed in the tumor interior, so that a decreasing radial concentration profile develops and drives the phenotype transitions that shape the growing tumor. Mechanically, the boundary rests on an elastic foundation provided by the host, whose stiffness varies with the polar angle such that $\mathbb{C}^{\external} = \mathbb{C}^{\external} ( \vartheta )$. Rigid-body motion is eliminated by pinning the center of the disk and supporting one boundary point on a roller, as indicated in Figure~\ref{fig:example_setup}. Fluid and solute exchange across $\partial \Oreference$ is modeled by permeable-membrane (Robin) conditions in which the normal fluxes are driven by the jump in the associated potentials. Accounting for the change of boundary area during growth, the reference normal fluxes on $\partial \Oreference$ read
\begin{subequations}
    \begin{align}
        \bJ_{\!\mathrm{f}} \cdot \bN &= \Js \bigl\lVert \bFs^{-\T} \bN \bigr\rVert \frac{L_{\mathrm{f}}}{\rho_{\mathrm{f}}} \Bigl( \mu_{\lambda,\mathrm{f}} - \mu_{\lambda,\mathrm{f}}^{\mathrm{ext}} - \tilde{\kappa}_{\mathrm{f}}^{\mathrm{o}} \bigl( \co - c_{\mathrm{f}}^{\mathrm{o,ext}} \bigr) - \tilde{\kappa}_{\mathrm{f}}^{\mathrm{w}} \bigl( \cw - c_{\mathrm{f}}^{\mathrm{w,ext}} \bigr) \Bigr) \,, \\
        \bQ_{\mathrm{f}}^{\mathrm{o}} \cdot \bN &= \Js \bigl\lVert \bFs^{-\T} \bN \bigr\rVert L_{\mathrm{f}}^{\mathrm{o}} \bigl( \tilde{\kappa}_{\mathrm{f}}^{\mathrm{o}} - \tilde{\kappa}_{\mathrm{f}}^{\mathrm{o},\mathrm{ext}} \bigr) \,,\\
        \bQ_{\mathrm{f}}^{\mathrm{w}} \cdot \bN &= \Js \bigl\lVert \bFs^{-\T} \bN \bigr\rVert L_{\mathrm{f}}^{\mathrm{w}} \bigl( \tilde{\kappa}_{\mathrm{f}}^{\mathrm{w}} - \tilde{\kappa}_{\mathrm{f}}^{\mathrm{w},\mathrm{ext}} \bigr) \,,
    \end{align}
\end{subequations}
with non-negative boundary permeabilities $L_{\mathrm{f}} = L_{\mathrm{f}} ( \vartheta )$, $L_{\mathrm{f}}^{\mathrm{o}} = L_{\mathrm{f}}^{\mathrm{o}} ( \vartheta )$, and $L_{\mathrm{f}}^{\mathrm{w}} = L_{\mathrm{f}}^{\mathrm{w}} ( \vartheta )$. The energetic state of the host is defined in analogy to the mixing energy \eqref{eq:tumor_psi_mix}, yielding the external fluid potential
\begin{align}
    \mu_{\lambda,\mathrm{f}}^{\external} = E_{\mathrm{f}} ( \log ( \phif^{\external} ) + 1 ) + \frac{A \phi_{\s}^{\external}}{c^{*}} c_{\mathrm{f}}^{\mathrm{o,ext}} + \frac{\epsilon^{\mathrm{o}}}{2} \left( \frac{c_{\mathrm{f}}^{\mathrm{o,ext}}}{c^{*}} \right)^{2} + \frac{\epsilon^{\mathrm{w}}}{2} \left( \frac{c_{\mathrm{f}}^{\mathrm{w,ext}}}{c^{*}} \right)^{2} + \lambda^{\external} \,,
\end{align}
and the corresponding weighted solute potentials
\begin{align}
    \tilde{\kappa}_{\mathrm{f}}^{\mathrm{o,ext}} = \frac{A \phi_{\s}^{\external}}{c^{*}} + \epsilon^{\mathrm{o}} \frac{c_{\mathrm{f}}^{\mathrm{o,ext}}}{( c^{*} )^{2}} \,,\qquad  \tilde{\kappa}_{\mathrm{f}}^{\mathrm{w,ext}} = \epsilon^{\mathrm{w}} \frac{c_{\mathrm{f}}^{\mathrm{w,ext}}}{( c^{*} )^{2}} \,.
\end{align}
Solutes are additionally advected across $\partial \Oreference$ by the transmembrane fluid flux $\bJ_{\!\mathrm{f}}$, for which the upwind concentration is used, that is, the interior value where fluid leaves the domain ($\bJ_{\!\mathrm{f}} \cdot \bN > 0$) and the reservoir value where it enters ($\bJ_{\!\mathrm{f}} \cdot \bN < 0$). The cellular constituents neither cross the boundary nor exert micro-tractions on it,
\begin{align}
    \bJ_{\!\mathrm{p}} \cdot \bN = \bJ_{\!\mathrm{h}} \cdot \bN = \bJ_{\!\mathrm{n}} \cdot \bN = 0 \,, \qquad
    \frac{\partial \Psi_{0}}{\partial \nablaX \phi_{\alpha}} \cdot \bN = 0 \,, \quad \alpha \in \{ \mathrm{p}, \mathrm{h}, \mathrm{n} \} \quad \text{on } \partial \Oreference \,.
\end{align}
The elastic support of the host tissue is described by a Robin condition that resists normal expansion of the boundary,
\begin{align}
    \bP \bN = - \mathbb{C}^{\external} ( \vartheta ) ( \bu \cdot \bN ) \bN \quad \text{on } \partial \Oreference \,.
\end{align}
Unless stated otherwise, all model parameters are treated as non-dimensional. The constant values used throughout the simulations are collected in Table~\ref{tab:tumor_parameters}. These values are chosen to be representative, rather than being fitted to a specific experimental data set. A quantitative parameter identification is left for future work.

\subsection{Unconfined growth}\label{sec:unconfined}
We first consider the reference case of unconfined growth. The boundary rests on a vanishing elastic support, $\mathbb{C}^{\external} = 0$, so that the tumor may expand freely, and the membrane permeabilities are spatially uniform, $L_{\mathrm{f}} = 10$ and $L_{\mathrm{f}}^{\mathrm{o}} = L_{\mathrm{f}}^{\mathrm{w}} = 250$. Under these conditions, the problem retains its rotational symmetry and the tumor grows radially, which allows all fields to be examined as radial profiles.

\begin{figure}[pos=htbp]
    \centering
    \includegraphics[width=0.7\textwidth]{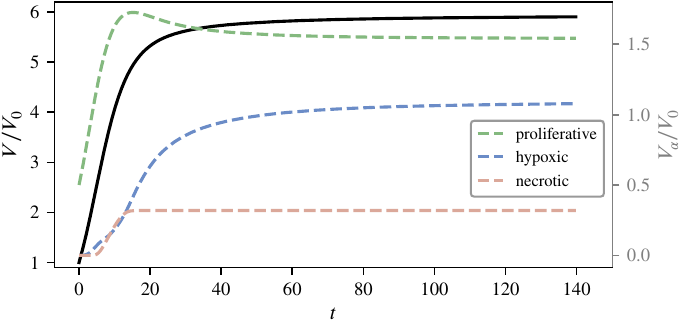}
    \caption{Unconfined growth history. Evolution of the relative tumor volume $V/V_{0}$, with $V_{0}$ the initial tumor volume, together with the per-phenotype volume ratios $V_{\!\alpha}/V_{0}$ as the tumor approaches mechanical equilibrium.}
    \label{fig:unconfined_history}
\end{figure}

The growth history is shown in Figure~\ref{fig:unconfined_history}. Starting from the homogeneous, fully proliferative initial state, the total tumor volume $V$ grows rapidly and then saturates as the tumor approaches equilibrium, reaching almost six times its initial value. The per-phenotype volume ratios $V_{\!\alpha}/V_{0}$ with reference volume $V_{0} = \pi R_{0}^{2}$ reveal the emergence of the characteristic phenotype layering: the proliferative volume overshoots and subsequently relaxes as interior cells turn hypoxic, the hypoxic volume rises monotonically toward a plateau, and a necrotic population forms once the innermost cells are depleted of oxygen. At equilibrium, all three populations are stationary, reflecting a balance between the nutrient supplied at the boundary and its consumption in the interior.

\begin{figure}[pos=htbp]
    \centering
    \includegraphics[width=0.95\textwidth]{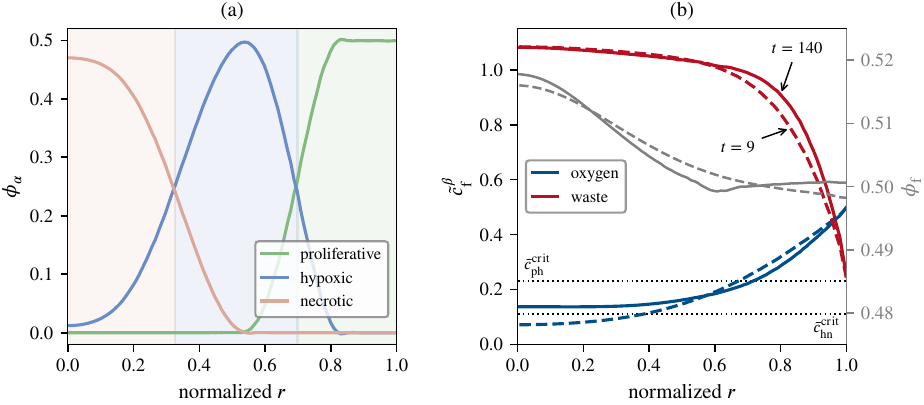}
    \caption{Radial profiles over the normalized radial position for unconfined growth. (a)~The phenotype volume fractions $\phi_{\alpha}$ at equilibrium, revealing a proliferative rim, a hypoxic shell, and a necrotic core (shaded). (b)~The fluid-borne oxygen and waste concentrations $\tilde{c}_{\mathrm{f}}^{\beta}$ at equilibrium ($t = 140$, solid) and at an earlier, still-growing state ($t = 9$, dashed). The core oxygen dips below $\bar{c}_{\mathrm{hn}}^{\mathrm{crit}}$ during growth and recovers above it near equilibrium, while the necrotic core persists. Fluid slightly accumulates in the tumor core.}
    \label{fig:unconfined_profiles}
\end{figure}

The radial structure of the tumor is reported in Figure~\ref{fig:unconfined_profiles} as a function of the normalized radial position, which permits a direct comparison of states of different tumor sizes. Panel~(a) shows the phenotype volume fractions $\phi_{\alpha}$ at equilibrium, which organize into three concentric zones: a well-oxygenated proliferative rim, an intermediate hypoxic shell, and a necrotic core. Panel~(b) explains this layering through the fluid-borne solute concentrations. The oxygen concentration increases from the poorly supplied core toward the boundary, whereas the metabolic waste concentration shows the opposite trend, peaking in the core. During rapid growth, the expanding, proliferation-rich tumor consumes oxygen faster than diffusion can supply it, so that the core oxygen concentration transiently falls below the critical value $\bar c_{\mathrm{hn}}^{\mathrm{crit}}$ and triggers the irreversible hypoxic–necrotic transition. As growth slows toward equilibrium, consumption decreases and the core oxygen concentration recovers to settle between the two thresholds. Having risen above $\bar{c}_{\mathrm{hn}}^{\mathrm{crit}}$, it switches off the irreversible hypoxic--necrotic transition and thereby freezes the necrotic core at the size reached during the transient, while remaining below $\bar{c}_{\mathrm{ph}}^{\mathrm{crit}}$ keeps the surrounding tissue hypoxic. The fluid volume fraction $\phi_{\mathrm{f}}$ slightly accumulates in the center of the tumor, leading to a swollen inner core.

\begin{figure}[pos=t]
    \centering
    \includegraphics[width=0.7\textwidth]{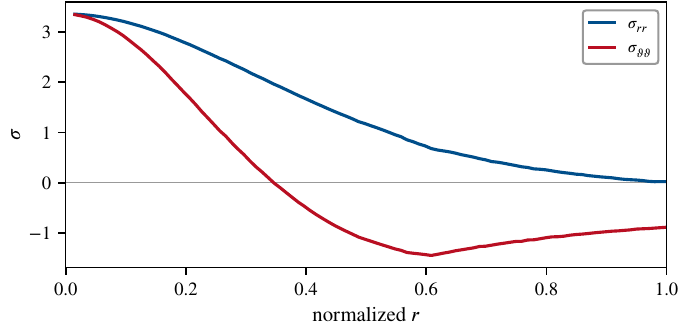}
    \caption{Growth-induced residual Cauchy stress $\bsigma$ at equilibrium over the normalized radial position for unconfined growth, with radial component $\sigma_{rr}$ and hoop component $\sigma_{\vartheta\vartheta}$, exhibiting a tensile core and a compressive rim.}
    \label{fig:unconfined_stress}
\end{figure}

Finally, Figure~\ref{fig:unconfined_stress} shows the growth-induced residual Cauchy stress at equilibrium. Since growth proceeds at different rates across the tumor, the locally grown material elements are geometrically incompatible, and the elastic deformation that restores a coherent, continuous body leaves the tumor stressed even in the absence of external loading. The radial component $\sigma_{rr}$ is tensile throughout and decays to zero at the traction-free boundary, whereas the hoop component $\sigma_{\vartheta\vartheta}$ is tensile in the core and turns compressive toward the rim. By rotational symmetry the two components coincide at the center. This state originates in the growth mismatch between the strongly proliferating rim and the quiescent interior: as the rim expands, it remains bonded to the weakly growing core and stretches it outward, placing the interior in radial and circumferential tension, while the rim itself, held back at its inner edge, is left in circumferential compression. The resulting tensile-core, compressive-rim pattern is the characteristic mechanical signature of differential growth and agrees with the residual stress states reported in the literature \citep{Ambrosi_2002, Olaranont_2025}.

\subsection{Confined growth}\label{sec:confined}
We next investigate how mechanical confinement by the surrounding host affects tumor growth. To describe a three-fold angular heterogeneity, we introduce the normalized modulation function
\begin{align}
    \mathcal{H} ( \vartheta ) \coloneq \frac{1}{16} ( 1 + \cos ( 3\vartheta ) )^{4} \,,
    \label{eq:threefold_modulation}
\end{align}
which varies between zero and one (see also the dashed lines in Figure~\ref{fig:example_setup}). To separate the influence of the magnitude and the spatial distribution of the elastic support, we compare a uniformly confined tumor, $\mathbb{C}^{\external} ( \vartheta ) = \mathbb{C}_{0}$, with a non-uniformly confined tumor subjected to the three-fold angular stiffness distribution $\mathbb{C}^{\external} ( \vartheta ) = \mathbb{C}_{0} \mathcal{H} ( \vartheta )$. We set $\mathbb{C}_{0} = 15$ in both cases and retain the uniform membrane permeabilities of Section~\ref{sec:unconfined}. The non-uniform support reaches the same maximum stiffness as the uniform support in three sectors, but vanishes in the intervening sectors. It therefore introduces a purely mechanical three-fold anisotropy into the otherwise rotationally symmetric problem.

\begin{figure}[pos=htbp]
    \centering
    \includegraphics[width=0.95\textwidth]{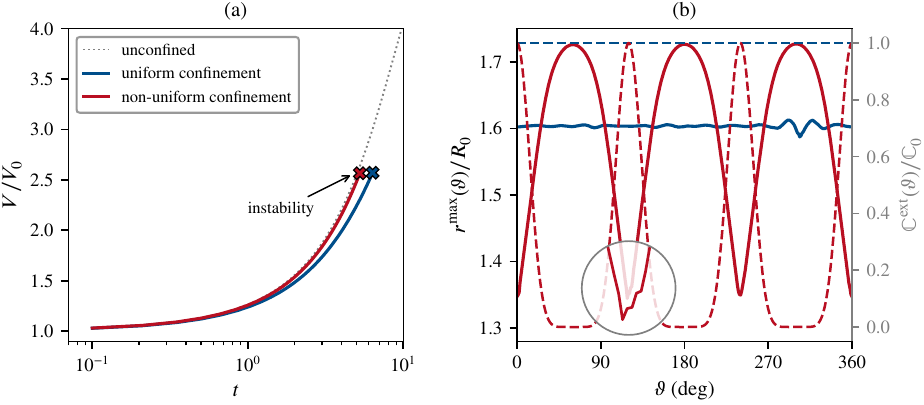}
    \caption{Growth and deformation under mechanical confinement. (a)~Evolution of the relative tumor volume $V/V_{0}$ for uniform and non-uniform confinement, together with the unconfined reference response. The crosses mark the onset of surface instabilities, at which the confined simulations are terminated. (b)~Normalized boundary radius $r^{\mathrm{max}} ( \vartheta )/R_{0}$ at the onset of the instabilities (solid, left axis) and normalized stiffness of the elastic support $\mathbb{C}^{\external} ( \vartheta )/\mathbb{C}_{0}$ (dashed, right axis).}
    \label{fig:buckling_history}
\end{figure}

Figure~\ref{fig:buckling_history} compares the growth histories and boundary shapes of the two confined tumors. For reference, the volume history of the unconfined tumor is included in Figure~\ref{fig:buckling_history}(a). All three responses are nearly identical during the initial growth stage, when the boundary displacements and the resulting foundation tractions are still small. As the tumor expands, confinement increasingly suppresses the global volume increase relative to the unconfined case. The uniformly confined tumor grows slightly more slowly than the non-uniformly confined tumor because it is supported along the entire boundary. Nevertheless, both confined tumors develop surface instabilities at nearly the same critical volume, $V^{\mathrm{crit}} \approx 2.6 V_{0}$. The non-uniformly confined tumor reaches this instability slightly earlier. We emphasize that surface instabilities, creasing, and folded morphologies of confined, differentially growing soft tissues, including tumor spheroids, have been documented both experimentally and theoretically \citep{BenAmar_2005, Li_2011, Dervaux_2011, Ciarletta_2013, Ziegler_2026}. However, since our finite element setup is not designed to handle the strong nonlinearities arising from such instabilities, the computations terminate at their onset.

The corresponding boundary profiles are shown in Figure~\ref{fig:buckling_history}(b). Under uniform confinement, the tumor remains almost circular with $r^{\mathrm{max}}/R_{0} \approx 1.6$ up to the instability, apart from a small localized perturbation associated with the emerging instabilities. By contrast, the non-uniformly confined tumor develops a pronounced three-fold shape. Its radius is maximal where the elastic support vanishes and minimal where $\mathbb{C}^{\external}$ reaches $\mathbb{C}_{0}$. The radius and stiffness profiles are therefore in antiphase: the tumor bulges into the compliant sectors and forms necks at the stiff sectors. This deformation pattern shows that the heterogeneous host redirects growth toward mechanically favorable directions even though the growth mobility itself is isotropic.

\begin{figure}[pos=htbp]
    \centering
    \includegraphics[width=0.95\textwidth]{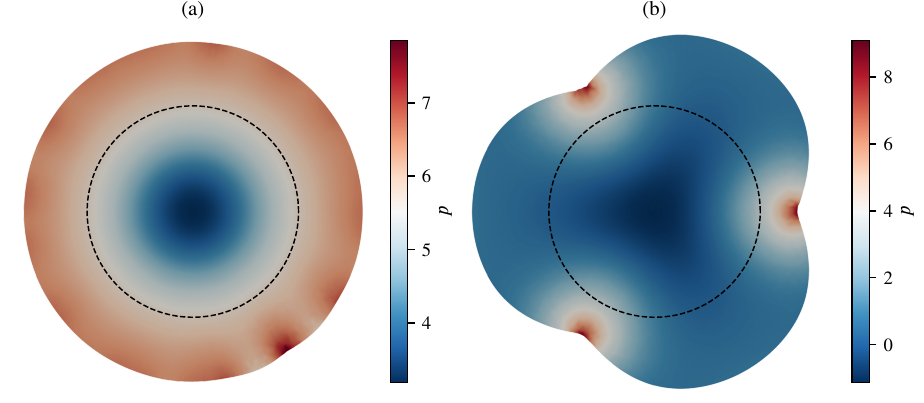}
    \caption{Hydrostatic pressure $p$ at the onset of instabilities under (a)~uniform and (b)~non-uniform mechanical confinement. Uniform confinement produces a broadly distributed increase in pressure toward the supported boundary and a localized maximum at the incipient instability. Non-uniform confinement instead produces three pressure concentrations at the stiff necks, while the tumor expands into the intervening compliant sectors. The dashed circles indicate the initial tumor boundary.}
    \label{fig:confined_fields}
\end{figure}

Figure~\ref{fig:confined_fields} shows how these different boundary constraints are reflected in the hydrostatic pressure
\begin{align}
    p \coloneq - \frac{1}{d} \tr ( \bsigma )
\end{align}
at the onset of instabilities. Uniform confinement generates a broadly distributed pressure that increases from the tumor interior toward the supported boundary, while the nearly circular shape is retained. The localized pressure maximum at the boundary coincides with the incipient instability perturbation that develops due to imperfections induced by the irregular finite element mesh. Under non-uniform confinement, the pressure remains substantially lower throughout most of the tumor, because large parts of the boundary can expand into the compliant sectors. Instead, pronounced pressure concentrations develop at the three stiff necks, where the host locally opposes growth. These concentrations identify the sites at which the mechanical constraint is transmitted most strongly into the tumor and at which instabilities initiate. Notably, the neck concentrations even exceed the peak pressure of the uniformly confined tumor, so that heterogeneous support does not merely relocate the pressure but raises its local maximum. Hence, the two confinement patterns lead to similar total volumes at instability but markedly different deformation and pressure fields: uniform support distributes the mechanical resistance over the boundary, whereas heterogeneous support concentrates it in isolated sectors and thereby organizes the growing tumor into a lobed configuration.

\subsection{Growth under heterogeneous boundary transport}\label{sec:transport}
In the final scenario, we break the rotational symmetry through heterogeneous transport rather than mechanical confinement. We remove the elastic support by setting $\mathbb{C}_{0} = 0$, so that the tumor boundary is mechanically free, and modulate all boundary permeabilities by the three-fold function $\mathcal{H}$ introduced in \eqref{eq:threefold_modulation}. Starting from the values of Section~\ref{sec:unconfined}, we prescribe $L_{\mathrm{f}} ( \vartheta ) = 10 \mathcal{H} ( \vartheta )$ and $L_{\mathrm{f}}^{\mathrm{o}} ( \vartheta ) = L_{\mathrm{f}}^{\mathrm{w}} ( \vartheta ) = 250 \mathcal{H} ( \vartheta )$. Fluid exchange, oxygen supply, and waste removal are therefore concentrated in three permeable sectors and vanish at the three intervening sectors. Since the mechanical boundary condition remains rotationally symmetric, any resulting anisotropy in the tumor shape originates from the coupling between heterogeneous transport, phenotype transitions, and growth. The computation is again terminated at the onset of surface instabilities, which occur at a critical volume of $V^{\mathrm{crit}} \approx 4.9 V_{0}$.

\begin{figure}[pos=htbp]
    \centering
    \includegraphics[width=0.95\textwidth]{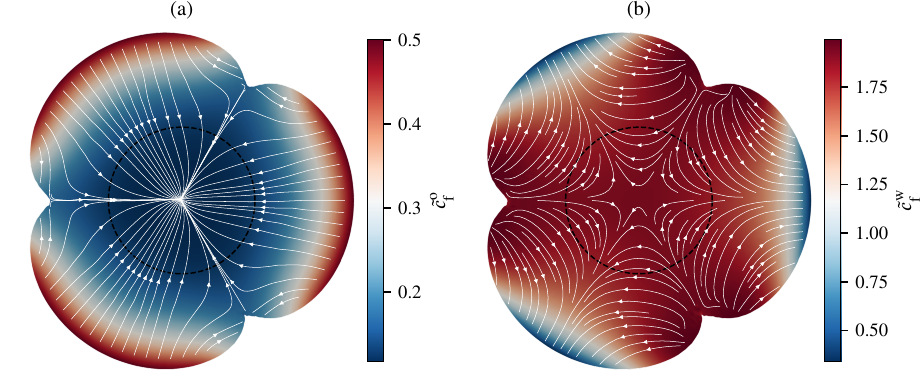}
    \caption{Fluid-borne solute fields under heterogeneous boundary transport, shown on the deformed tumor configuration. (a)~Oxygen concentration $\tilde{c}_{\mathrm{f}}^{\mathrm{o}}$ and streamlines of the oxygen flux $\bQ_{\mathrm{f}}^{\mathrm{o}}$. Oxygen enters through the three permeable sectors and is transported toward the depleted tumor interior. (b)~Waste concentration $\tilde{c}_{\mathrm{f}}^{\mathrm{w}}$ and streamlines of the waste flux $\bQ_{\mathrm{f}}^{\mathrm{w}}$. Waste produced in the interior is transported outward and removed through the same permeable sectors. The dashed circles indicate the initial tumor boundary.}
    \label{fig:nutrient_fields}
\end{figure}

The corresponding solute fields on the deformed configuration are shown in Figure~\ref{fig:nutrient_fields}. Oxygen enters through the permeable sectors and, as indicated by the inward-directed streamlines of $\bQ_{\mathrm{f}}^{\mathrm{o}}$ in Figure~\ref{fig:nutrient_fields}(a), is transported toward the interior where it is consumed, so that its concentration is highest at the well-supplied lobes and decreases toward the core. Metabolic waste exhibits the complementary pattern in Figure~\ref{fig:nutrient_fields}(b): produced throughout the active tissue, it accumulates in the interior and is transported outward, so that its concentration is lowest at the permeable outlets and elevated in the interior and toward the poorly permeable boundary. The permeable sectors thus act simultaneously as nutrient inlets and waste outlets.

\begin{figure}[pos=htbp]
    \centering
    \includegraphics[width=0.95\textwidth]{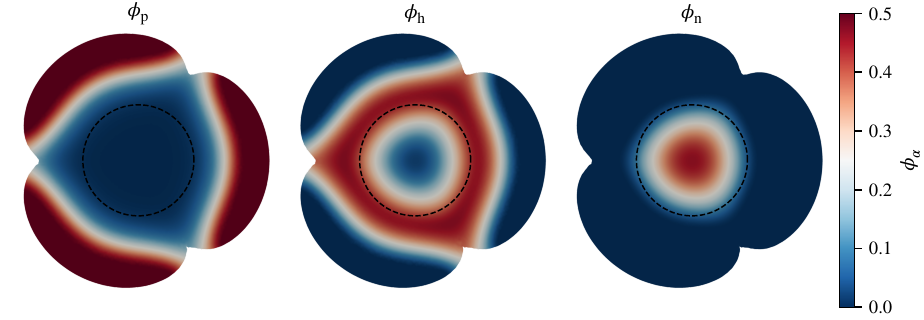}
    \caption{Cellular phenotype volume fractions under heterogeneous boundary transport, shown on the deformed tumor configuration. From left to right: proliferative cells $\phip$, hypoxic cells $\phih$, and necrotic cells $\phin$. The proliferative phenotype forms along the well-supplied lobes, the hypoxic phenotype forms a three-fold intermediate shell, and the necrotic phenotype is concentrated in the oxygen-depleted core. The dashed circles indicate the initial tumor boundary.}
    \label{fig:cell_fields}
\end{figure}

The heterogeneous nutrient supply directly determines the spatial organization of the cellular phenotypes, as shown in Figure~\ref{fig:cell_fields}. Proliferative cells occupy the well-oxygenated outer region and form a thick rim along the three supplied lobes. This rim narrows strongly at the poorly permeable necks, where the oxygen concentration is lower. The hypoxic phenotype forms an intermediate shell that separates the proliferative rim from the necrotic core. In contrast to the nearly concentric layering observed for unconfined growth, this shell inherits the three-fold nutrient pattern and extends farther toward the boundary in the poorly supplied sectors. The necrotic phenotype remains localized in an approximately circular central core, where the oxygen concentration reaches its minimum.

Because the growth mobility \eqref{eq:tumor_growth_mobility} degenerates with both the proliferative volume fraction and the available oxygen, growth is sustained primarily in the well-supplied sectors. The tumor consequently expands preferentially in the directions of maximum permeability and develops three pronounced lobes, while the impermeable sectors remain recessed. The shape modulation is therefore in phase with the boundary permeability. This behavior contrasts with the non-uniformly confined case of Section~\ref{sec:confined}, where the tumor bulges in the directions of minimum mechanical stiffness. The two examples demonstrate that similar lobed morphologies may emerge from distinct mechanisms: heterogeneous mechanical resistance redirects an otherwise uniform growth process, whereas heterogeneous transport first reorganizes the phenotype distribution and thereby localizes the growth itself.

\section{Conclusion and outlook}\label{sec:conclusion}
In this work, we developed a thermodynamically consistent continuum framework for the finite growth of fully saturated multi-constituent mixtures containing an arbitrary number of dilute dissolved solutes. Starting from constituent-wise balance laws in a solid-skeleton-based description, we derived the admissible constitutive structure from a free-energy dissipation principle, which yields process-wise dissipation inequalities that constrain the closures for mass exchange, constituent and solute transport, chemical reactions, and growth. The principal novelty of the framework is the coupling between growth-induced volume creation and constituent mass production, expressed through volume accumulation fractions that distribute the newly created volume among the constituents while preserving saturation, thereby generalizing classical single-solid growth closures to genuine multi-constituent mixtures. To the best of our knowledge, this is the first framework to unite finite growth kinematics and fully saturated multi-constituent mixture theory with dilute solute transport within a single thermodynamically consistent setting. We cast the resulting model in a total Lagrangian mixed weak form amenable to standard finite element solvers and specialized the general theory to a four-constituent, two-solute model of avascular tumor growth that couples nutrient transport, waste production, phenotype transitions between proliferative, hypoxic, and necrotic cells, volume growth, elastic deformation, and growth-induced residual stress. The computational examples illustrate the coupled mechanisms captured by this formulation. Unconfined growth produces the characteristic phenotype layering and a tensile-core, compressive-rim residual stress state. Mechanical confinement limits growth, triggers surface instabilities, and localizes pressure under non-uniform support. Heterogeneous boundary transport produces comparable lobing through spatially localized nutrient supply. These results show that similar morphologies may arise from distinct mechanical or biochemical mechanisms.

The framework opens several future research avenues. On the modeling side, the isotropic growth mobility could be generalized to a structure-dependent or anisotropic tensor following the direction of tissue fibers, and the solid skeleton could be endowed with inelastic effects such as viscoelastic behavior. Moreover, a homeostatic stress state acting as a growth limiter might be introduced, and growth-induced surface instabilities such as buckling and creasing require further investigation. On the biological side, the avascular tumor model provides a basis for extensions toward vascularized and therapy-coupled growth, while systematic parameter identification against experimental data, together with dedicated experimental validation of the predicted morphologies and residual stress states, would enable quantitative predictions. Finally, the numerical implementation leaves room for adaptive discretizations.

\printcredits

\section*{Data availability}
The results can be reproduced using only the information presented in the manuscript. All data and the developed Python code can be made available upon request from the corresponding author.

\section*{Declaration of competing interest}
The authors declare that they have no known competing financial interests or personal relationships that could have appeared to influence the work reported in this paper.

\section*{Funding}
M.F.P. ten Eikelder acknowledges support from the German Research Foundation (Deutsche Forschungsgemeinschaft DFG), project number 566600860. Open access funding enabled and organized by Projekt DEAL.

\section*{Declaration of generative AI and AI-assisted technologies in the manuscript preparation process}
During the preparation of this work, the authors used GPT-5.6 Sol and Claude Opus 4.8 in order to improve the grammar, language, and readability of the manuscript. Moreover, Figure~\ref{fig:tumor_spheroid} was partly generated using GPT-5.6 Sol. After using these tools, the authors reviewed and edited the content as needed and take full responsibility for the content of the published article.

\appendix
\section{Symmetry of the host-switching mobility}\label{sec:host_switching_symmetry}
In the absence of local solute reactions, the concentration source terms \eqref{eq:split_reaction_source} reduce to
\begin{align}
    r_{\alpha}^{\beta} = \sum_{\gamma=1}^{N} \hat{r}^{\mathrm{sw},\beta}_{\alpha \gamma} \,.
\end{align}
Inserting this expression together with the linear exchange law \eqref{eq:solute_host_switching} into the solute-reaction dissipation inequality \eqref{eq:constraint_concentration_exchange} yields the host-switching contribution
\begin{align}
    - \sumalpha \sum_{\gamma = 1}^{N} \sumbeta \ell_{\!\alpha \gamma}^{\beta}  \tilde{\kappa}_{\alpha}^{\beta} \bigl( \tilde{\kappa}_{\gamma}^{\beta} - \tilde{\kappa}_{\alpha}^{\beta} \bigr) \geq 0 \,.
\end{align}
Relabeling the summation indices $\alpha \leftrightarrow \gamma$ leaves this expression unchanged and gives the equivalent representation
\begin{align}
    - \sum_{\gamma = 1}^{N} \sumalpha \sumbeta \ell_{\!\gamma \alpha}^{\beta} \tilde{\kappa}_{\gamma}^{\beta} \bigl( \tilde{\kappa}_{\alpha}^{\beta} - \tilde{\kappa}_{\gamma}^{\beta} \bigr) \geq 0 \,.
\end{align}
Averaging these two inequalities and invoking the symmetry $\ell_{\!\alpha\gamma}^{\beta} = \ell_{\!\gamma\alpha}^{\beta}$, we obtain
\begin{align}
    - \sumalpha \sum_{\gamma = 1}^{N} \sumbeta \frac{1}{2} \ell_{\!\alpha \gamma}^{\beta} \Bigl( \tilde{\kappa}_{\alpha}^{\beta} \bigl( \tilde{\kappa}_{\gamma}^{\beta} - \tilde{\kappa}_{\alpha}^{\beta} \bigr) + \tilde{\kappa}_{\gamma}^{\beta} \bigl( \tilde{\kappa}_{\alpha}^{\beta} - \tilde{\kappa}_{\gamma}^{\beta} \bigr) \Bigr) = \sumalpha \sum_{\gamma = 1}^{N} \sumbeta \frac{1}{2} \ell_{\!\alpha \gamma}^{\beta} \bigl( \tilde{\kappa}_{\gamma}^{\beta} - \tilde{\kappa}_{\alpha}^{\beta} \bigr)^{2} \geq 0 \,.
\end{align}
The symmetry of the mobility therefore renders the host-switching contribution to the dissipation a sum of squares, which is non-negative, and hence thermodynamically admissible.

Furthermore, for a given pair of host constituents $\alpha$ and $\gamma$, the linear exchange law \eqref{eq:solute_host_switching} provides the two switching rates
\begin{align}
    \hat{r}^{\mathrm{sw},\beta}_{\alpha \gamma} = \ell_{\!\alpha \gamma}^{\beta} \bigl( \tilde{\kappa}_{\gamma}^{\beta} - \tilde{\kappa}_{\alpha}^{\beta} \bigr) \,,\qquad \hat{r}^{\mathrm{sw},\beta}_{\gamma \alpha} = \ell_{\!\gamma \alpha}^{\beta} \bigl( \tilde{\kappa}_{\alpha}^{\beta} - \tilde{\kappa}_{\gamma}^{\beta} \bigr) \,.
\end{align}
Using the symmetry $\ell_{\!\alpha\gamma}^{\beta} = \ell_{\!\gamma\alpha}^{\beta}$ once more, these rates are skew-symmetric under exchange of the two host constituents,
\begin{align}
    \hat{r}^{\mathrm{sw},\beta}_{\alpha \gamma} + \hat{r}^{\mathrm{sw},\beta}_{\gamma \alpha} = 0 \,.
\end{align}
Consequently, the amount of solute $\beta$ transferred out of host $\alpha$ is exactly received by host $\gamma$, so that the switching process merely redistributes each solute among the constituents without creating or destroying it.

\section{Evaluation of the constitutive quantities for the tumor growth model}\label{sec:tumor_constitutive}
By inserting the free-energy density \eqref{eq:tumor_psi_mech}--\eqref{eq:tumor_psi_int} into the general constitutive relations of Section~\ref{sec:total_lagrangian_formulation}, we obtain the explicit constitutive quantities of the tumor model, namely the chemical potentials, stresses, fluxes, and exchange and reaction rates.

\paragraph{Chemical potentials.}
The mechanical part of the free-energy density \eqref{eq:tumor_psi_mech} is independent of the volume fractions and does not contribute to the saturation-augmented chemical potentials of the constituents $\mu_{\lambda,\alpha} = \Js^{-1} \mu_{0\lambda,\alpha}$. Instead, the bulk contributions stem from the mixing \eqref{eq:tumor_psi_mix} and fitness \eqref{eq:tumor_psi_fit} energies, while the interfacial energy \eqref{eq:tumor_psi_int} produces non-local divergence terms. Carrying out the differentiation yields, for the three cellular phenotypes,
\begin{subequations}
    \begin{align}
        \mu_{\lambda,\mathrm{p}} &= E_{\mathrm{s}} ( \log \phi_{\mathrm{s}} + 1 ) + \frac{A}{c^{*}} \tilde{c}_{\mathrm{f}}^{\mathrm{o}} - \frac{B}{c^{*}} \bigl( \tilde{c}_{\mathrm{f}}^{\mathrm{o}} - \bar{c}_{\mathrm{ph}}^{\mathrm{crit}} \bigr) + \lambda - \frac{\gamma}{\Js} \Div \bigl( \Js \bC_{\!\mathrm{s}}^{-1} \nablaX \phip \bigr) \,, \\
        \mu_{\lambda,\mathrm{h}} &= E_{\mathrm{s}} ( \log \phi_{\mathrm{s}} + 1 ) + \frac{A}{c^{*}} \tilde{c}_{\mathrm{f}}^{\mathrm{o}} + \lambda - \frac{\gamma}{\Js} \Div \bigl( \Js \bC_{\!\mathrm{s}}^{-1} \nablaX \phih \bigr) \,, \\
        \mu_{\lambda,\mathrm{n}} &= E_{\mathrm{s}} ( \log \phi_{\mathrm{s}} + 1 ) + \frac{A}{c^{*}} \tilde{c}_{\mathrm{f}}^{\mathrm{o}} + \frac{B}{c^{*}} \bigl( \tilde{c}_{\mathrm{f}}^{\mathrm{o}} - \bar{c}_{\mathrm{hn}}^{\mathrm{crit}} \bigr) + \lambda - \frac{\gamma}{\Js} \Div \bigl( \Js \bC_{\!\mathrm{s}}^{-1} \nablaX \phin \bigr) \,.
    \end{align}
\end{subequations}
For the fluid, which carries no interfacial energy, we obtain
\begin{align}
    \mu_{\lambda,\mathrm{f}} = E_{\mathrm{f}} ( \log \phif + 1 ) + \frac{( A - B ) \phip + A \phih + ( A + B ) \phin}{c^{*}} \co + \frac{\epsilon^{\mathrm{o}}}{2} \left( \frac{\co}{c^{*}} \right)^{2} + \frac{\epsilon^{\mathrm{w}}}{2} \left( \frac{\cw}{c^{*}} \right)^{2} + \lambda \,.
    \label{eq:chemical_potential_fluid}
\end{align}
Similarly, transforming the Lagrangian weighted chemical potentials of the solutes to the current configuration yields
\begin{subequations}
    \begin{align}
        \tilde{\kappa}_{\mathrm{f}}^{\mathrm{o}} &= \frac{( A - B ) \phip + A \phih + ( A + B ) \phin}{c^{*}} + \epsilon^{\mathrm{o}} \frac{\co}{( c^{*} )^{2}} \,,\\
        \tilde{\kappa}_{\mathrm{f}}^{\mathrm{w}} &= \epsilon^{\mathrm{w}} \frac{\cw}{( c^{*} )^{2}}  \,,
    \end{align}
        \label{eq:chemical_potentials_solutes}%
\end{subequations}
respectively.

\paragraph{Stress measures.}
The stress response is obtained by inserting the free-energy density \eqref{eq:tumor_psi_mech}--\eqref{eq:tumor_psi_int} into the first Piola--Kirchhoff stress \eqref{eq:PK1_stress}, which requires the derivatives of the individual free-energy contributions with respect to the elastic deformation gradient. The mechanical part yields the standard neo-Hookean expression, whereas the mixing and fitness parts depend on $\bFe$ only through the Jacobian $\Js$ by virtue of $\partial \Js / \partial \bFe = \Js \bFe^{-\T}$. The interfacial part also depends on $\bFe$ through $\bC_{\!\mathrm{s}}^{-1}$ appearing in the gradient norm. For the bulk parts of the free-energy we thus obtain
\begin{align}
    \frac{\partial \Psi_{0}^{\mathrm{mech}}}{\partial \bFe} = G \bigl( \bFe - \bFe^{-\T} \bigr) \,,\qquad \frac{\partial \Psi_{0}^{\mathrm{mix}}}{\partial \bFe} = \Psi_{0}^{\mathrm{mix}} \bFe^{-\T} \,,\qquad \frac{\partial \Psi_{0}^{\mathrm{fit}}}{\partial \bFe} = \Psi_{0}^{\mathrm{fit}} \bFe^{-\T} \,,
\end{align}
and differentiating the interface energy gives
\begin{align}
    \frac{\partial \Psi_{0}^{\mathrm{int}}}{\partial \bFe} = \Psi_{0}^{\mathrm{int}} \bFe^{-\T} - \sum_{\alpha \in \{ \mathrm{p}, \mathrm{h}, \mathrm{n} \}} \Js \gamma ( \nablax \phi_{\alpha} \otimes \nablax \phi_{\alpha} ) \bFe^{-\T} \,.
\end{align}
Together with the identity $\bFs^{-\T} = \bFe^{-\T} \bFg^{-\T}$, the first Piola--Kirchhoff stress is then assembled as
\begin{align}
    \bP = G \bigl( \bFe \bFe^{\T} - \bI \bigr) \bFs^{-\T} + \Js \bvarsigma \bFs^{-\T} - \left( \Psi_{0}^{\mathrm{mech}} + \sum_{\alpha \in \{ \mathrm{p}, \mathrm{h}, \mathrm{n}, \mathrm{f} \}} \Js \phi_{\alpha} \mu_{\lambda,\alpha} \right) \bFs^{-\T} \,.
\end{align}
Here, the mechanical free energy $\Psi_{0}^{\mathrm{mech}}$ and the chemical pressures $\phi_{\alpha} \mu_{\lambda,\alpha}$, which carry the saturation multiplier $\lambda$, enter as isotropic, pressure-like contributions, whereas the free-energy pressure and the interfacial gradient terms combine into the symmetric Korteweg stress
\begin{align}
    \bvarsigma = \frac{\Psi_{0}}{\Js} \bI - \gamma \sum_{\alpha \in \{ \mathrm{p}, \mathrm{h}, \mathrm{n} \}} \nablax \phi_{\alpha} \otimes \nablax \phi_{\alpha} \,.
\end{align}
The corresponding Cauchy stress follows as
\begin{align}
    \bsigma = \frac{G}{\Js} \bigl( \bFe \bFe^{\T} - \bI \bigr) + \bvarsigma - \left( \frac{\Psi_{0}^{\mathrm{mech}}}{\Js} + \sum_{\alpha \in \{ \mathrm{p}, \mathrm{h}, \mathrm{n}, \mathrm{f} \}} \phi_{\alpha} \mu_{\lambda, \alpha} \right) \bI \,.
\end{align}
Finally, the Eshelby stress that drives growth evaluates to
\begin{align}
    \bSigma = \bigl( \Psi_{0} + \Psi_{0}^{\mathrm{mech}} \bigr)  \bI - G ( \bCe - \bI ) - \Js \bFe^{\T} \bvarsigma \bFe^{-\T} \,.
\end{align}

\paragraph{Fluid transport and solute diffusion.}
Inserting the chemical potentials and weighted solute potentials \eqref{eq:chemical_potential_fluid} and \eqref{eq:chemical_potentials_solutes} into the constitutive flux laws of Section~\ref{sec:total_lagrangian_formulation} yields the explicit constituent flux
\begin{align}
    \bJ_{\!\mathrm{f}} = - \frac{\bM_{0,\mathrm{ff}}}{\rho_{\mathrm{f}}} \left( \frac{E_{\mathrm{f}}}{\phif} \nablaX \phif + \frac{\co}{c^{*}} \left( ( A - B ) \nablaX \phip + A \nablaX \phih + ( A + B ) \nablaX \phin  \right) + \nablaX \lambda \right) \,,
\end{align}
and the solute diffusive fluxes
\begin{subequations}
    \begin{align}
        \bQ_{\mathrm{f}}^{\mathrm{o}} &= - \frac{\bK_{0,\mathrm{ff}}^{\mathrm{oo}}}{c^{*}} \left( ( A - B ) \nablaX \phip + A \nablaX \phih + ( A + B ) \nablaX \phin + \frac{\epsilon^{\mathrm{o}}}{c^{*}} \nablaX \co \right) \,,\\
        \bQ_{\mathrm{f}}^{\mathrm{w}} &= - \frac{\bK_{0,\mathrm{ff}}^{\mathrm{ww}}}{c^{*}} \frac{\epsilon^{\mathrm{w}}}{c^{*}} \nablaX \cw \,.
    \end{align}
\end{subequations}

\paragraph{Cell phenotype transitions and oxygen consumption.}
With the forward affinity kinetics \eqref{eq:constitutive_choice_mass_exchange}, the transition affinities \eqref{eq:tumor_affinities} yield the nominal cellular production terms
\begin{subequations}
    \begin{align}
        Z_{\mathrm{p}}^{\exchange} &= m_{0,\mathrm{h} \to \mathrm{p}} \bigl\langle \mathcal{A}_{0,\mathrm{h} \to \mathrm{p}} \bigr\rangle_{\!+} - m_{0,\mathrm{p} \to \mathrm{h}} \bigl\langle \mathcal{A}_{0, \mathrm{p} \to \mathrm{h}} \bigr\rangle_{\!+} \,,\\
        Z_{\mathrm{h}}^{\exchange} &= m_{0,\mathrm{p} \to \mathrm{h}} \bigl\langle \mathcal{A}_{0, \mathrm{p} \to \mathrm{h}} \bigr\rangle_{\!+} - m_{0,\mathrm{h} \to \mathrm{p}} \bigl\langle \mathcal{A}_{0,\mathrm{h} \to \mathrm{p}} \bigr\rangle_{\!+} - m_{0,\mathrm{h} \to \mathrm{n}} \bigl\langle \mathcal{A}_{0, \mathrm{h} \to \mathrm{n}} \bigr\rangle_{\!+} \,,\\
        Z_{\mathrm{n}}^{\exchange} &= m_{0,\mathrm{h} \to \mathrm{n}} \bigl\langle \mathcal{A}_{0, \mathrm{h} \to \mathrm{n}} \bigr\rangle_{\!+} \,.
    \end{align}
\end{subequations}
The fluid-borne oxygen-to-waste reaction is driven by the affinity
\begin{align}
    \mathcal{B}_{0,\mathrm{f}}^{\mathrm{o} \to \mathrm{w}} = \frac{( A - B ) \phip + A \phih + ( A + B ) \phin}{c^{*}} + \frac{\epsilon^{\mathrm{o}} \co}{( c^{*} )^{2}} - \frac{\mathsf{M}^{\mathrm{o}}}{\mathsf{M}^{\mathrm{w}}} \frac{\epsilon^{\mathrm{w}} \cw}{( c^{*} )^{2}} \,,
    \label{eq:tumor_reaction_affinity}
\end{align}
which yields the solute source terms
\begin{align}
    R_{\mathrm{f}}^{\mathrm{o}} = -k_{0,\mathrm{f}}^{\mathrm{o} \to \mathrm{w}} \bigl\langle \mathcal{B}_{0,\mathrm{f}}^{\mathrm{o} \to \mathrm{w}} \bigr\rangle_{\!+} \,,\qquad
    R_{\mathrm{f}}^{\mathrm{w}} = k_{0,\mathrm{f}}^{\mathrm{o} \to \mathrm{w}} \frac{\mathsf{M}^{\mathrm{o}}}{\mathsf{M}^{\mathrm{w}}} \bigl\langle \mathcal{B}_{0,\mathrm{f}}^{\mathrm{o} \to \mathrm{w}} \bigr\rangle_{\!+} \,.
\end{align}

\section{Regularization of entropic mixing energy}\label{sec:polynomial_approximation}
The entropic mixing contributions $f ( \phi_{\alpha} ) \coloneq \phi_{\alpha} \log ( \phi_{\alpha} )$ in the mixing energy \eqref{eq:tumor_psi_mix} are defined only for positive volume fractions, and their derivatives $f^{\prime} ( \phi_{\alpha} ) = \log ( \phi_{\alpha} ) + 1$, which enter the chemical potentials of Appendix~\ref{sec:tumor_constitutive}, become singular as $\phi_{\alpha} \to 0$. During the solution of the nonlinear, discrete problem, the volume fractions may leave the admissible interval $[ 0, 1 ]$, where the logarithm and its derivative are undefined or unbounded, which frequently causes the iteration to break down.

\begin{figure}[pos=htbp]
    \centering
    \includegraphics[width=0.95\textwidth]{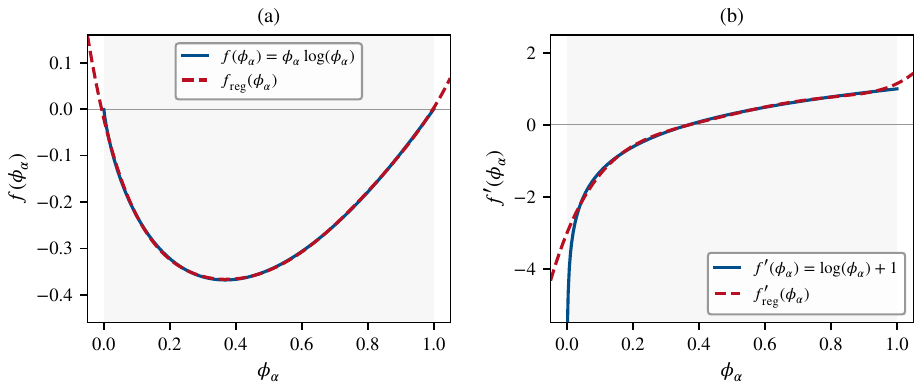}
    \caption{Polynomial regularization of the entropic mixing term. (a)~The exact term $f ( \phi_{\alpha} )$ and its degree-six polynomial approximation $f_{\mathrm{reg}} ( \phi_{\alpha} )$ on the admissible interval $\phi_{\alpha} \in [ 0, 1 ]$. (b)~The corresponding derivatives: the exact derivative $f^{\prime} ( \phi_{\alpha} )$ diverges as $\phi_{\alpha} \to 0$, whereas the regularized derivative $f_{\mathrm{reg}}^{\prime} ( \phi_{\alpha} )$ remains finite, which is essential for the robustness of the finite element solution.}
    \label{fig:regularization}
\end{figure}

To avoid this, we replace the entropic term by a global polynomial approximation of degree six \citep{Eikelder_2026},
\begin{align}
    f ( \phi_{\alpha} ) = \phi_{\alpha} \log ( \phi_{\alpha} ) \approx f_{\mathrm{reg}} ( \phi_{\alpha} ) \coloneq \sum_{i = 0}^{6} a_{i} \phi_{\alpha}^{i} \,,
\end{align}
with coefficients obtained from a least-squares fit on $[ 0, 1 ]$,
\begin{align*}
    &a_{0} = -0.0209 \,,\quad &&a_{1} = - 2.9992 \,,\quad &&a_{2} = 11.2314 \,,\quad &&a_{3} = - 24.9257 \,,\\
    &a_{4} = 34.2319 \,,\quad &&a_{5} = - 24.6192 \,,\quad &&a_{6} = 7.1045 \,.
\end{align*}
Both $f_{\mathrm{reg}}$ and its derivative $f_{\mathrm{reg}}^{\prime} ( \phi ) = \sum_{i = 1}^{6} i a_{i} \phi^{i-1}$ are smooth and finite on all of $\mathbb{R}$, so that the mixing energy and the associated chemical potentials remain well-defined even when a volume fraction temporarily overshoots $[ 0, 1 ]$. Accordingly, wherever the entropic derivative $\log ( \phi_{\alpha} ) + 1$ appears in the chemical potentials of Appendix~\ref{sec:tumor_constitutive}, it is evaluated through $f_{\mathrm{reg}}^{\prime}$ in the numerical implementation. Figure~\ref{fig:regularization} compares the exact term and its derivative with the polynomial approximation. On the admissible interval the two are in close agreement, while the approximation removes the singularity at $\phi = 0$ and extends smoothly beyond $[ 0, 1 ]$.

\bibliographystyle{cas-model2-names}

\bibliography{references}

\end{document}